\documentclass[preprint]{aastex}
\def\hinvMpc{$h^{-1}\,{\rm Mpc}$ }
\def\hmpc{{h^{-1}\,{\rm Mpc}}}

\slugcomment{Submitted to ApJ}

\shorttitle{Halo Populations}

\shortauthors{Zheng, Tinker, Weinberg, \& Berlind}

\begin{document}

\title{Do Distinct Cosmological Models Predict Degenerate Halo 
Populations?}
\author{Zheng Zheng$^1$, Jeremy L. Tinker$^1$, 
David H. Weinberg$^1$, and Andreas A. Berlind$^2$}
\affil{{}$^1$ Department of Astronomy, The Ohio State University,
Columbus, OH 43210, USA}
\affil{{}$^2$ Department of Astronomy \& Astrophysics, The University 
of Chicago, Chicago, IL 60637, USA}
\email{zhengz,tinker,dhw@astronomy.ohio-state.edu}
\email{aberlind@oddjob.uchicago.edu}

\begin{abstract}
Using cosmological N-body simulations, we investigate the influence of 
the matter density parameter $\Omega_m$ and the linear theory power 
spectrum $P(k)$ on statistical properties of the dark matter halo 
population --- the mass function $n(M)$, two-point correlation function 
$\xi(r)$, and pairwise velocity statistics $v_{12}(r)$ and 
$\sigma_{12}(r)$.  For fixed linear theory $P(k)$, the effect of 
changing $\Omega_m$ is simple: the halo mass scale $M_*$ shifts in 
proportion to $\Omega_m$, pairwise velocities (at fixed $M/M_*$) are 
proportional to $\Omega_m^{0.6}$, and halo clustering at fixed $M/M_*$ 
is unchanged.  If one simultaneously changes the power spectrum amplitude 
$\sigma_8$ to maintain the ``cluster normalization'' condition 
$\sigma_8\Omega_m^{0.5}=$const, then $n(M)$ stays approximately constant 
near $M\sim 5\times 10^{14}h^{-1}M_\odot$, and halo clustering and 
pairwise velocities are similar at fixed $M$.  However, the shape
of $n(M)$ changes, with a decrease of $\Omega_m$ from 0.3 to 0.2 
producing a $\sim 30\%$ drop in the number of low mass halos.  One can 
preserve the shape of $n(M)$ over a large dynamic range by changing the 
spectral tilt $n_s$ or shape parameter $\Gamma$, but the required changes 
are substantial --- e.g., masking a decrease of $\Omega_m$ from 0.3 to 
0.2 requires $\Delta n_s\approx 0.3$ or $\Delta\Gamma\approx 0.15$.  
These changes to $P(k)$ significantly alter the halo clustering and halo 
velocities.  The sensitivity of the dark halo population to cosmological 
model parameters has encouraging implications for efforts to constrain 
cosmology and galaxy bias with observed galaxy clustering, since the 
predicted changes in the halo population cannot easily be masked by 
altering the way that galaxies occupy halos.  A shift in $\Omega_m$ alone 
would be detected by any dynamically sensitive clustering statistic; a 
cluster normalized change to $\sigma_8$ and $\Omega_m$ would require a 
change in galaxy occupation as a function of $M/M_*$, which would alter 
galaxy clustering; and a simultaneous change to $P(k)$ that preserves the 
halo mass function would change the clustering of the halos themselves.
\end{abstract}


\section{Introduction}

In cosmological models dominated by cold dark matter (CDM) and an 
unclustered energy component (such as a cosmological constant), 
gravitational instability of primordial density fluctuations produces a 
population of dark matter halos, each in approximate virial equilibrium.  
Depending on their mass, these halos may host individual galaxies, galaxy 
groups, or rich galaxy clusters.  Provided that one focuses on systems of 
overdensity $\rho/\bar{\rho}\sim 200$, and regards higher density 
structures as subsystems within their parent halos, the halo population 
itself is insensitive to the gas pressure forces that influence the 
sub-dominant baryon component.  In this paper, we investigate whether 
changes to cosmological parameters --- specifically the matter density 
parameter $\Omega_m$ and parameters that describe the shape and amplitude 
of the primordial power spectrum --- always produce measurable changes in
the dark halo population, or whether two models with different 
combinations of these parameters can give rise to halo populations that 
are effectively indistinguishable.

Our interest in this question is spurred by recent developments in the
theory of biased galaxy formation.  The uncertain relation between
galaxies and dark matter is the primary limitation in testing cosmological
models against observations of galaxy clustering.  The ``halo occupation
distribution'' (HOD) characterizes this relation statistically in terms
of the probability distribution $P(N|M)$ that a halo of virial mass $M$
contains $N$ galaxies of a specified type, together with prescriptions
that specify the relative spatial and velocity distributions of galaxies
and dark matter within these halos.  Numerous recent papers have shown
that the HOD framework is a powerful tool for analytic and numerical
calculations of clustering statistics, for modeling observed clustering,
and for characterizing the results of semi-analytic or numerical studies
of galaxy formation (e.g., \citealt{kauffmann97,jing98b,kauffmann99,
benson00,ma00,peacock00,seljak00,berlind02,bullock01,marinoni02,
scoccimarro01,yoshikawa01,white01}).  In particular, \cite{peacock00}, 
\cite{marinoni02}, and \cite{berlind02} have argued that the HOD can be 
determined empirically from observed galaxy clustering, given an assumed 
cosmological model that determines the mass function and spatial and 
velocity clustering of the dark halo population.  Empirical determinations 
of the HOD can provide insight into the physics of galaxy formation, and 
they may sharpen the ability of large scale structure studies to test 
cosmological models, since a model with an incorrect dark halo population 
may be unable to match the data for any choice of HOD (see 
\citealt{berlind02} for further discussion).

Suppose that we find a combination of cosmology and HOD that reproduces
all aspects of observed galaxy clustering.  Can we infer that the 
cosmology and the derived HOD are both correct?  If there is another 
cosmological model that predicts an indistinguishable halo population, 
then the answer is clearly no, since the combination of this alternative 
cosmology with the same HOD would predict identical galaxy clustering.
These cosmological models would be degenerate in the sense that they
could not be distinguished by galaxy clustering data without relying on
a predictive theory of galaxy formation (which might yield different HODs 
for the two cosmologies).  Cosmological models that predict 
distinguishable halo populations could still be degenerate in this
sense, if changes to the HOD can mask the differences in the halo 
populations; we will speculate on this point in \S 4, but we reserve
a quantitative examination of it to future work.

We focus our analyses on CDM cosmological models with Gaussian initial 
density fluctuation fields.  Motivated by cosmic microwave background 
measurements \citep{netterfield02,pryke02}, we restrict our attention 
to spatially flat models with a cosmological constant or $\Omega_m=1$,
though we demonstrate in passing that galaxy clustering data at $z=0$
probably cannot distinguish a flat model from an open model with the
same $\Omega_m$. The parameters that define our cosmological models are:
the matter density parameter $\Omega_m$; the normalization $\sigma_8$ of 
the power spectrum $P(k)$, which is the r.m.s. fluctuation of the linear 
density field filtered with a top-hat filter of radius 8\hinvMpc; and 
the shape of $P(k)$, which we characterize by the spectral index $n_s$ 
of the inflationary power spectrum and the shape parameter $\Gamma$ of 
the matter transfer function (see, e.g., \citealt{efstathiou92}). 
Although existing data impose constraints on these parameters, 
individually or in combination, here we allow each to vary independently 
over a fairly broad range, so that we can isolate the physical effects 
of the matter density, the amplitude of mass fluctuations, and the shape 
of the power spectrum on the resulting halo population.  We concentrate 
entirely on the halo populations at $z=0$, since this is where galaxy 
clustering data will be good enough to allow empirical HOD determinations 
in the near future.

Using $N$-body simulations described in \S\ref{sec:sims}, we measure (in 
\S\ref{sec:results}) mass functions, two-point correlation functions, 
mean pairwise radial velocities, and pairwise velocity dispersions of 
dark matter halos. First, we consider models that have the same initial 
power spectrum $P(k)$ but different matter density parameter $\Omega_m$. 
We then move to cluster normalized models in which $\sigma_8$ is changed 
to compensate the change in $\Omega_m$, so that the amplitude of the halo
mass function is kept approximately fixed at a cluster scale. Finally, we 
investigate models in which both the amplitude and the shape of $P(k)$ 
are changed in order to match the amplitude and slope of the halo mass 
function at the cluster scale.  We review our results and briefly discuss 
their implications for cosmological tests in \S\ref{sec:discussion}.
Our results overlap previous numerical studies of the halo mass function
and halo clustering (recent examples include \citealt{jing98,governato99,
colberg00,jenkins01}), but they differ in the examination of controlled
parameter sequences rather than specific cosmological models. Also, while
some of these studies have focused on the large scale correlations of
cluster mass halos, we devote considerable attention to the lower masses
and smaller spatial scales that are important for non-linear galaxy 
clustering.  Throughout the paper, we adopt a Hubble constant 
$H_0 = 100 h\; {\rm km\, s^{-1}\, Mpc^{-1}} = 
70\; {\rm km\, s^{-1}\, Mpc^{-1}} $ \citep{freedman01}.

\section{Numerical Methods and Tests}
\label{sec:sims}

In this work, statistics are mainly based on simulations run with a 
Particle-Mesh (PM) $N$-body code \citep{park90}. We run PM simulations 
with different combinations of density and power spectrum parameters. 
Each simulation follows the evolution of 200$^3$ particles in a periodic 
cube 200\hinvMpc on a side, with a 400$^3$ mesh to compute the 
gravitational force starting at $z=19$ and advancing to $z=0$ in 40 
equal steps of expansion factor $a$. The mass of each particle is 
$2.78 \times 10^{11} \Omega_m h^{-1}M_\odot$. We adopt the 
parameterization of \cite{efstathiou92} for the fluctuation power 
spectrum $P(k)$.  We divide our simulations into three categories 
corresponding to parameter changes in the matter density and the 
normalization and the shape of the initial fluctuation power spectrum. 
For each category, four cosmological models are investigated with 
$(\Omega_m,\Omega_\Lambda)=$(0.2,0.8), (0.3,0.7), (0.4,0.6) and 
(1.0,0.0). We generally regard the (0.3,0.7) model as the central one 
for comparison.  In the central model, the spectral index is $n_s=1.0$, the 
spectral shape parameter is $\Gamma=0.20$, and the r.m.s. mass fluctuation 
within spheres of radius 8\hinvMpc is $\sigma_8=0.9$, which is consistent 
with the observed abundance of clusters \citep{eke96}. For each model, 
we generate four independent realizations, and we use the dispersion among 
these realizations to estimate the statistical error bars on our 
clustering measures associated with the finite simulation volume. The 
total volume is comparable to that expected for the Sloan Digital Sky 
Survey \citep{york00} out to its median redshift 
$z_{\rm med} \approx 0.1$. We use a FoF algorithm \citep{davis85} to 
identify the dark matter halos, and, unless explicitly stated otherwise, 
the linking length is 0.2 times the mean inter-particle separation.
With this linking length, the FoF algorithm picks out structures of 
typical mass overdensity $\rho/\bar{\rho} \sim 200$ \citep{davis85}.
 
The advantage of a PM code is that it is relatively cheap to run multiple 
simulations with a large dynamic range in mass. The limitation of a 
PM code is its moderate force resolution, but this limitation is not 
too serious for our purposes, since we only identify halos and do not 
attempt to measure their internal structure. We perform two tests 
to make sure that the PM resolution is adequate for our purpose. First, 
we make a comparison between PM simulations and simulations run with the
GADGET tree code \citep{springel01}. Second, we perform a self-similar 
scaling test for the PM code.

GADGET is a publicly available $N$-body code that calculates particle 
accelerations by the hierarchical tree method of \cite{barnes86}. Its 
advantages over the PM algorithm are its spatial and temporal 
adaptability.  The gravitational softening length, $\epsilon$, can be 
set significantly smaller than the initial particle grid spacing to 
increase force resolution, though if $\epsilon$ is too small 2-body 
relaxation effects will begin to influence the halo population. 
Particles have individual timesteps, which can vary continuously to 
ensure accurate time integration given the adopted force resolution. 
GADGET has fully periodic boundary conditions for comoving integrations 
via the Ewald summation method \citep{hernquist91}.

We have compared the results of two simulations with 100$^3$ particles
in a 100\hinvMpc box, one evolved with the PM code using a 200$^3$ mesh 
(thus having equal resolution to the larger simulations on which our 
results will be based), and one evolved with GADGET using 
$\epsilon=0.3h^{-1}$Mpc. The initial conditions of the two simulations 
were identical to each other and used the parameters of the central 
model described above. Slices through these simulations are plotted for 
comparison in Figure~\ref{fig:pgslice}.  The structures formed in the 
two simulations are very similar at the level of detail discernible in 
this plot. The halo mass functions of the two simulations, also shown 
in Figure~\ref{fig:pgslice}, trace each other nearly exactly. The PM 
simulation produces slightly ($\sim 10\%$) fewer halos below 
$10^{13} h^{-1}M_\odot$, which is to be expected given its lower force 
resolution; the diameter of an overdensity 200 sphere of 
$10^{13} h^{-1}M_\odot$ is 1.04$\hmpc$, or about two PM grid cells. 
When the mesh size on this PM simulation was increased to 300$^3$, the 
resulting mass function was consistent with the GADGET simulation even 
on these small mass scales.

In addition to the halo mass function, the quantities important to our
analysis are the correlation functions for both the mass distributions
and the halo populations, which are shown in Figure~\ref{fig:pgxicmp}. 
As with the halo mass functions, the matter correlation functions for
the PM and GADGET simulations match very well. They begin to deviate at 
separations smaller than $\sim 0.5 h^{-1}$Mpc, or under one mesh cell. 
The halo correlation functions produced by the two methods are also 
consistent.  Larger simulations and multiple realizations of each model 
are efficiently produced with the PM code, which takes up to an order 
of magnitude less computing time at the force resolution adopted here.

We next perform a self-similar scaling test for the PM code.  Assuming 
an Einstein-de Sitter universe ($\Omega_m=1$), we use a pure power-law 
power spectrum $P(k) \propto k^{-n_s}$ with $n_s=-1$ (and no transfer 
function modulation) for the initial density fluctuations of the 
scale-free model. A 400$^3$ force mesh is used to follow the 
gravitational evolution of 200$^3$ particles. The normalization of the 
power spectrum is set so that the nonlinear mass $M_*$ at the 
final output corresponds to 400 particles. We define $M_*$ by the 
condition $\sigma(M_*)=\delta_c=1.69$, where $\sigma(M_*)$ is the 
r.m.s. linear mass fluctuation on mass scale $M_*$. The r.m.s. 
fluctuation is $\sigma=1.0$ at about 1/26 of the box size.  If we 
identify this scale as 8\hinvMpc, the implied size of the periodic 
simulation cube is then 207.62\hinvMpc at $z=0$.  Earlier outputs can 
be identified with higher redshifts or with larger volumes at $z=0$. 
In the language of the redshift description, the simulation is evolved 
for 40 time steps from $z=19$, with 4 outputs (at redshifts $z=$1.00, 
0.60, 0.25, and 0.00), when the corresponding $M_*$ values are 50, 100, 
200, and 400 particles, respectively. The following results are 
averaged over four independent realizations of the simulation, and the 
error bars represent the dispersion among the four realizations divided 
by $\sqrt{3}$ to yield the error on the mean. 

\citet{efstathiou88} performed the first study on scale-free 
cosmological models using $N$-body simulations and showed that many 
quantities in scale-free models have well-defined scaling behaviors.  
In an Einstein-de Sitter scale-free model, the cumulative halo mass 
functions at redshifts $z_1$ and $z_2$ have the scaling behavior
\begin{equation}
\label{eqn:mscaling}
N(M>M_2; z_2)=\left(\frac{1+z_2}{1+z_1}\right)^{6/(3+n_s)} 
  N(M>M_1; z_1), ~~~ M_1=\left(\frac{1+z_2}{1+z_1}\right)^{6/(3+n_s)}M_2.
\end{equation}
The left panel of Figure~\ref{fig:scalemf} shows the cumulative halo 
mass function at different redshifts in our scale-free simulations 
(solid curves).  The dotted curves in Figure~\ref{fig:scalemf} are
derived by scaling the mass function at $z=0$ according to the above
scaling rule. The scaling properties shown in the figure are generally 
very good. The high mass end of the mass function drops systematically
below the self-similar scaling as $M_*$ increases, a likely result of
the finite simulation volume, which supresses the amplitude of 
fluctuations on the largest scales. Comparison of the first and last 
outputs suggests that this effect becomes noticeable above 
$M \sim 2\times10^{15} h^{-1}M_\odot$ at $z=0$ (for our adopted 
$\sigma_8=1.0$ scaling), or a mass scale $\sim$ 0.1\% of the total 
simulation mass (10\% in length scale).

The $N$-point correlation functions $\xi(r)$ are expected to have a 
similarity transformation with the variable $s=r(1+z)^{2/(3+n_s)}$ for
an Einstein-de Sitter universe \citep{efstathiou88}. The scaling rule 
for the correlation functions at redshifts $z_1$ and $z_2$ is then
\begin{equation}
\label{eqn:xiscaling}
\xi(r_2; z_2)=\xi(r_1; z_1), ~~~ r_1=\left(\frac{1+z_2}{1+z_1}\right)^{2/(3+n_s)}r_2.
\end{equation}
From Figure~\ref{fig:scalemf} and Figure~\ref{fig:scalexi}, it is clear 
that mass correlation functions and halo correlation functions at 
different values of $M/M_*$ obey the scaling rule quite well.  There 
are some noticeable departures from self-similar scaling in the 
$M_*/2$ panel at the two earliest output times (especially the 
earliest), when $M_*/2$ corresponds to 25 and 50 particles, 
respectively.

The clustering of halos is biased relative to that of the matter. 
The halo bias factor $b_h$ can be defined via 
$\xi_h(r)=b_h^2\xi_m(r)$, where $\xi_h$ and $\xi_m$ are the 
two-point correlation functions of halos and matter, respectively.
The relation between the clustering of halos and that of the matter has 
been extensively studied based on analytical models and numerical
simulations (e.g., \citealt{cole89,kashlinsky91,mo96,jing98,jing99,
porciani99,sheth99a,sheth99,sheth01}; and references therein).
\citet{mo96} give an analytic formula for the bias factor of halos
of a given mass based on an extended Press-Schechter analysis 
\citep{press74}. This formula is quite accurate for halos with 
$M>M_*$. \citet{jing98} empirically modifies the formula to also fit 
the $N$-body simulation results for halos of lower mass, 
\begin{equation}
b_h(M,z)=\left(\frac{1}{2\nu^4}+1\right)^{(0.06-0.02n_s)}
\left(1+\frac{\nu^2-1}{\delta_{c,0}}\right), ~~~ \nu=\delta_c(z)/\sigma(M),
\end{equation}
where $\sigma(M)$ is the r.m.s. fluctuation of the mass density 
at a mass scale $M$, $\delta_c(z)$ is the threshold density contrast 
for collapse of a homogeneous spherical overdense region at redshift 
$z$, and $\delta_{c,0}\approx 1.69$ for an Einstein-de Sitter universe. 
For $n_s=-1$ here, $\nu=(M/M_*)^{2/3}$.  We derive the square of the 
bias factor numerically as a function of $M/M_*$ at different 
redshifts by averaging $\xi_h(r)/\xi_m(r)$ over comoving separations 
between 5\hinvMpc and 30\hinvMpc. In this range, the ratio is almost 
constant. We compare the computed bias factor with that given by 
Jing's fitting formula in Figure~\ref{fig:scalebias}. Two conclusions 
can be drawn immediately from the comparison.  First, since bias 
factors at different redshifts overlap with each other when plotted 
as a function of $M/M_*$, the simulation displays the correct scaling 
behavior. Second, the bias factors agree well with Jing's fitting 
formula.  At higher mass, the computed bias is somewhat lower than 
that given by Jing's formula. In fact, the same trend can be found 
in Jing's plot (see his Figure~2). We attribute this slight deviation 
in part to a sample volume effect and in part to systematic errors in 
the approximate analytic formula given by \citet{mo96} (see 
\citealt{sheth01}).  

Both the GADGET simulation test and the scale-free model test assure 
that the numerical artifacts in the PM simulations have almost 
negligible effect on the statistics we care about in this paper. For 
example, the mass scales that we are concerned with are generally 
higher than the level where the PM and GADGET halo mass functions 
begin to deviate slightly.  Moreover, the correlation functions are 
measured at scales $\gtrsim$ 0.5 mesh cells, where the PM and GADGET 
results are consistent. The PM simulations reproduce the expected 
scaling behaviors in the scale-free model, which also demonstrates that 
we can measure masses of halos and their spatial distribution correctly 
based on these simulations. The scaling tests and code comparison 
suggest that there are some inaccuracies when the halo mass is lower 
than 50 particles or more than 0.1\% of the total mass in the 
simulation, but even in these regimes there should be little impact on 
our conclusions because we compare different cosmologies evolved with 
the same code and numerical parameters. The PM code is therefore an 
ideal tool for our purpose in this paper.

\section{Comparison of Halo Populations}
\label{sec:results}

\subsection{Changing $\Omega_m$ with $\sigma_8$, $n_s$, and $\Gamma$ Fixed}

We start with a simple case in which we fix the power spectrum (as 
linearly evolved to $z=0$, with $n_s=1.0$, $\sigma_8=0.9$, 
$\Gamma=0.20$) and change $\Omega_m$.  In linear theory, the evolved 
fluctuations in these models are identical, and this independence of 
$\Omega_m$ holds in the Zel'dovich approximation \citep{zeldovich70} 
as well as its extension, the adhesion approximation, as emphasized 
by \citet{weinberg90}. In fact, it holds remarkably well into the 
fully nonlinear regime, as shown in the top and middle panels of 
Figure~\ref{fig:slice}, where slices of particle distributions at 
$z=0$ for models with different $\Omega_m$ but identical initial 
conditions are compared. These slices are taken from GADGET 
simulations that evolve the particle distribution from $z=19$. 
Particle distributions of three spatially flat models with different 
values of $\Omega_m$ and one open model with $\Omega_m=0.2$ are 
extremely similar to each other (though particle velocities
are higher for higher $\Omega_m$).
\citet{nusser98} extend the Zel'dovich/adhesion analytic explanation
for this similarity of evolved structure by showing that
the equations of motion of a collisionless gravitating 
system of particles in an expanding universe can be put in a form with 
almost no dependence on $\Omega_m$ and $\Omega_\Lambda$,
if the linear perturbation growth factor is used as the time variable.
The evolutionary histories of the four models in Figure~\ref{fig:slice}
are different, but this difference is captured almost entirely by
the dependence of the linear growth factor on redshift.
The bottom two panels demonstrate this point, showing the particle 
distributions of the flat $\Omega_m=1.0$ and $\Omega_m=0.2$ models
at the redshifts when the linear growth factor is 0.5, redshifts
$z=1.0$ and $z=1.74$, respectively (we define the growth factor
to be unity at $z=0$.)
In most comparisons of N-body simulations with different $\Omega_m$,
the amplitude and/or shape of $P(k)$ is adjusted in concert with
$\Omega_m$.  Such adjustments are well motivated by observational
and theoretical considerations, but they mask the fact that the
influence of $\Omega_m$ at fixed $P(k)$ is extremely simple.
Earlier examples of matched comparisons like those in Figure~\ref{fig:slice}
include Figure~1 of \citet{davis85} and Figure~5 of \cite{nusser98}.

Figure~\ref{fig:slice} suggests that halo mass functions for different 
models in this case should be nearly the same when measured as a 
function of particle number, i.e., mass divided by $\Omega_m$. This 
expectation is confirmed in Figure~\ref{fig:case1mf}, which shows mass 
functions (from PM simulations) of four spatially flat models.  In the 
left panel, all four (cumulative) mass functions have a similar shape 
and only shift horizontally relative to each other.  When the halo 
masses are divided by $\Omega_m$, the four curves become nearly 
identical and overlap with the $\Omega_m=1$ curve. To allow a closer 
inspection, in the right panel we plot the ratio of the scaled mass 
functions to that calculated using Jenkins et al.'s (2001) fitting 
formula for the central model.  The maximum relative difference 
between the four scaled mass functions is only about 10\%. 

The 10\% residual differences reflect slight changes in the 
characteristic densities and profiles of halos in different 
cosmologies. For Figure~\ref{fig:case1mf}, we have identified halos with 
our standard FoF linking length parameter $b=0.2$, which picks out 
structures with mean overdensity $\rho/\bar{\rho} \sim 200$ 
(\citealt{davis85, lacey93}), close to the value 
$\rho/\bar{\rho} \sim 178$ predicted for the post-virialization
overdensity in the spherical collapse model for an $\Omega_m=1$ 
universe (\citealt{gunn72,peacock99}). If we instead use $b=0.16$, 
roughly doubling the overdensity threshold 
($0.2^3/0.16^3 \approx 1.95$), then halo masses drop by $\sim$20\% as 
previously linked particles are unlinked, but the level of agreement 
among different cosmological models stay the same.  Another fairly 
common procedure is to scale $b^{-3}$ in proportion to the virial 
overdensity predicted by the spherical collapse model, thus using a
different $b$ for each cosmological model. If we adopt this approach, 
taking the virial overdensities from \citet{eke96}, then the 
discrepancy of the scaled mass functions reverses sign and rises in 
amplitude to a maximum $\sim 20\%$. We conclude that the effect of a 
pure $\Omega_m$ change is described by the simple scaling of halo 
masses to an accuracy $\sim 10\%$ for halos defined at fixed 
overdensity, that the discrepancies reflect the expected cosmology 
dependence of the characteristic virial overdensity, but that the 
magnitude of the cosmology correction predicted by the spherical 
collapse model is too large. \citet{jenkins01} also
find better agreement among halo mass functions of different 
cosmological models for halos defined at constant overdensity 
rather than overdensities scaled according to the spherical collapse 
model.  Henceforth we will use the fixed linking length $b=0.2$ for 
all models, but our results would not be substantially different if 
we used $b=0.16$ or scaled $b$ with cosmology.

The mass correlation functions $\xi_m(r)$ for these four models are 
identical at large scales and gradually depart from each other at 
small scales ($r < 1$\hinvMpc), where the higher overdensities for 
lower $\Omega_m$ boost $\xi_m(r)$. We calculate the two-point 
correlation function for halos of mass $M$ by counting pairs of halos 
with masses in the range ($M/\sqrt{2}$, $\sqrt{2}M$). 
Figure~\ref{fig:case1xi} shows halo correlation functions $\xi_h(r)$ 
for four values of $M/M_*$.  In each case, the four $\Omega_m$ models 
are indistinguishable from each other, as we would expect, 
since $M_*$ scales with $\Omega_m$, and comparisons at fixed $M/M_*$ 
therefore remove any dependence on $\Omega_m$.  
The difference in $\xi_m(r)$ arises from differences in spatial
structure within the halos themselves (which are difficult to
discern in Figure~\ref{fig:slice}).
We also calculate the 
halo bias factor from the mass and halo correlation functions (not 
shown). When plotted as a function of $M/M_*$, the bias factors of 
the four models are almost the same, as expected, and they agree 
with Jing's (1998) fitting formula to a similar degree as seen in 
Figure~\ref{fig:scalebias}. The halo correlation function plummets at 
small $r$ because two halo centers cannot be separated by less than 
the sum of their virial radii. Halos for different $\Omega_m$ have 
the same virial radius at a given $M/M_*$ (in simulation terms, the 
scaling of $M_*$ with $\Omega_m$ comes from the particle mass, not a 
change in halo size), so even this exclusion signature is the same in 
different models.
 
Although mass functions and spatial clustering of halos are nearly 
identical as a function of $M/M_*$, different values of $\Omega_m$ 
produce different velocity fields. Figure~\ref{fig:case1ve} shows the 
mean pairwise radial (inward) velocity, 
$v_{12} \equiv - {\bf (v_1-v_2) \cdot (r_1-r_2) / |r_1-r_2|}$,  
and the pairwise velocity dispersion, 
$\sigma_{12} \equiv \langle v_{12}^2 \rangle - \langle v_{12} \rangle^2$, 
at $M=M_*$ and $M=8M_*$. Here ${\bf v}_i$ and ${\bf r}_i$ ($i=1,2$) 
are the velocities and positions of a halo pair, and the average 
is over all halo pairs with separations around $r$. Both $v_{12}(r)$ 
and $\sigma_{12}(r)$ drop as $\Omega_m$ drops. The mean pairwise radial 
velocity $v_{12}(r)$ increases for massive halos and increases much 
faster for pairs with small separations, a sign of the non-linear 
infall velocities induced by high mass halos.  The velocity dispersion 
$\sigma_{12}(r)$ continues rising out to large separations, and its 
amplitude at large scales seems to be nearly independent of halo mass.  
Linear perturbation theory predicts a relation between the peculiar 
velocity and matter density fields,
\begin{equation}
{\bf v(x)} \propto f(\Omega_m) \int \delta({\bf x}) 
\frac{({\bf x^\prime - x})}{|{\bf x^\prime - x}|^3} d^3x^\prime,
\end{equation}
where $\delta({\bf x})$ is the mass density contrast and $f(\Omega_m)
\approx \Omega_m^{0.6}$ (see, e.g., \citealt{peebles93}). In the four 
cosmological models considered here, the mass density contrasts 
$\delta({\bf x})$ should be the same, so linear velocity fields simply 
scale with $\Omega_m^{0.6}$. The dotted curves in 
Figure~\ref{fig:case1ve} show the effect of dividing $v_{12}(r)$ and 
$\sigma_{12}(r)$ by $\Omega_m^{0.6}$; the agreement of these curves 
with each other and with the $\Omega_m=1$ curve demonstrates that the 
impact of a pure $\Omega_m$ change on halo peculiar velocities is well 
described by the linear theory scaling even on non-linear scales.

Figure~\ref{fig:slice} suggests that the agreement of appropriately
scaled halo mass functions, correlation functions, and pairwise
velocity statistics should hold at other epochs for which the linear
growth factors are equal.  Though we do not show the results here,
we have verified explicitly that the matching of scaled statistical
measures in Figures~\ref{fig:case1mf}, \ref{fig:case1xi}, and~\ref{fig:case1ve}
holds equally well at the redshifts when the four models
have growth factors of 0.5 ($z=1.0$, 1.36, 1.50, and 1.74 for
$\Omega_m=1.0,$ 0.4, 0.3, 0.2).  Thus, for models with the same 
$P(k)$ shape and the same $\sigma_8$ at $z=0$, the influence of 
$\Omega_m$ and $\Omega_\Lambda$ on the evolutionary history of the 
halo population is entirely encoded in the linear growth factor.

\subsection{Cluster-normalized Models with Fixed $P(k)$ Shape}

Changing $\Omega_m$ on its own preserves the shape of the halo mass 
function but shifts the mass scale in proportion to $\Omega_m$. 
This scaling implies that the threshold volume for collapse is 
independent of $\Omega_m$, given the same power spectrum.  If we 
require that the four cosmological models that differ in $\Omega_m$ 
have halo mass functions that match in amplitude at some physical mass 
scale (rather than $M/M_*$), we need to adjust the power spectrum in a 
way that builds more massive halos in low-$\Omega_m$ models and fewer 
massive halos in high-$\Omega_m$ models. In other words, we need a 
larger threshold volume for collapse in a low-$\Omega_m$ model, so 
that its nonlinear mass $M_*$ increases. Therefore, if the shape of 
the power spectrum is kept unchanged, a higher $\sigma_8$ is necessary 
in a lower $\Omega_m$ model in order to compensate for the lower mean 
mass density.

Putting this idea in quantitative form, \citet{white93} argued that 
the observed abundances of massive clusters of galaxies impose the 
constraint $\sigma_8 \approx 0.57\Omega_m^{-0.56}$, with little 
dependence on the assumed shape of $P(k)$. Many authors have 
revisited, refined and recalibrated this ``cluster normalization" 
constraint (see \citealt{pierpaoli01} and references therein). One 
widely used formulation is that of \citet{eke96}, who found 
$\sigma_8=(0.52\pm0.04)\Omega_m^{-0.52+0.13\Omega_m}$
for a spatially flat universe. Inspired by these results, we now 
consider a sequence of cluster normalized models in which we keep 
the shape of $P(k)$ fixed but scale the amplitude as 
$\sigma_8=0.9(\Omega_m/0.3)^{-0.5}$.

Figure~\ref{fig:case2mf} shows cumulative halo mass functions for 
these cluster normalized models.  Mass functions of models with 
$\Omega_m=$0.2, 0.3, and 0.4 match in amplitude at a cluster scale 
($M \sim 5\times10^{14} h^{-1}M_\odot$). The mass function of the 
$\Omega_m=1$ model crosses the others at a somewhat lower mass scale,
$M \sim 2 \times 10^{14} h^{-1}M_\odot$. The discrepancy in matched 
mass scale just shows that the $\sigma_8 \propto \Omega_m^{-0.5}$ 
scaling does not hold all the way to $\Omega_m=1$, and the 
\citet{eke96} formula indeed implies that we should adopt a somewhat 
higher $\sigma_8$ for $\Omega_m=1$. However, the crucial result for
our purposes is that the mass functions of these cluster 
normalized models have systematically different shapes, and there is 
no value of $\sigma_8$ we could choose that would make the 
$\Omega_m=1$ model match the others at more than a single mass scale. 
Over the smaller range $\Omega_m=0.2-0.4$, the cluster normalization 
condition keeps the halo mass functions reasonably well aligned over 
about a decade in mass, but by $M \sim 10^{13} h^{-1}M_\odot$ the 
halo space densities of the $\Omega_m=0.2$ and 0.4 models differ from 
those of the central model by $\sim 30\%$, and from each other by 
nearly a factor of two.

The reason that cluster normalization does not preserve the shape 
of the halo mass function is that raising $\sigma_8$ increases $M_*$ 
but simultaneously drives down the space density of $M_*$ clusters 
by increasing the scale of non-linearity. For models with  the same 
$P(k)$ shape and the same value of $M_*$, the space density of halos 
at fixed $M/M_*$ is proportional to $\Omega_m$ (see equation~(7) 
below). The mass functions of cluster normalized models with 
different $\Omega_m$ must therefore cross at different values of 
$M/M_*$, and because the shapes remain the same as a function of 
$M/M_*$, they cannot stay aligned as a function of $M$.

In these cluster normalized models, the amplitude of $\xi_m(r)$ 
increases as $\Omega_m$ decreases, as shown in the upper left panel 
of Figure~\ref{fig:case2xi}.  At large scales, where the mass 
density field is still linear, $\xi_m(r)$ is just proportional to 
the amplitude of its Fourier transform, the linear power spectrum, 
and hence to $\sigma_8^2$.  The roughly constant logarithmic offset 
in Figure~\ref{fig:case2xi} shows that 
this multiplicative scaling holds approximately even into the 
strongly non-linear regime. However, the remaining panels of 
Figure~\ref{fig:case2xi} show that the halo correlation functions 
are very similar at a fixed physical mass.  Qualitatively, this 
similarity makes sense, since a given mass scale corresponds to a 
higher value of $M/M_*$ when $\Omega_m$ is higher, and the 
correspondingly higher halo bias tends to compensate for the lower 
mass clustering amplitude. However, it is impressive just how well
this cancellation between mass clustering and halo bias works 
quantitatively. The halo exclusion signature does shift to larger 
scale for lower $\Omega_m$ because of the larger virial radii 
needed to compensate the lower mass density.

Figure~\ref{fig:case2ve} shows that the mean pairwise velocity and
pairwise dispersion increase with $\Omega_m$ in this cluster 
normalized sequence, but the dependence is weak, especially at 
large separations.  As in Figure~\ref{fig:case1ve}, more massive 
halos have higher pairwise velocities, especially at small 
separations. The weak dependence on $\Omega_m$ at large scales can 
be understood with the help of equation~(4). In linear theory, 
velocity is proportional to $f(\Omega_m) \sigma_8$, which, when 
combined with the cluster normalization condition, reduces to 
approximately $\Omega_m^{0.1}$. Non-linear evolution at small 
separations amplifies the difference between models, but only 
slightly.

\subsection{Matching Mass Functions}

By varying the amplitude of the power spectrum with $\Omega_m$, 
we can match the amplitudes of the cumulative halo mass functions 
at a given mass scale. However, this combination of parameter 
changes does not guarantee a match between the shapes of the mass 
functions at that mass scale.  In order to match both the amplitude 
and the shape of the halo mass functions, we must adjust the 
relative amplitudes of fluctuations at different scales, or, in 
other words, the shape of $P(k)$.  We therefore need to 
determine the appropriate combinations of parameters ($\sigma_8$ 
and parameters related to the shape of the power spectrum) for each 
$\Omega_m$ model so that all models produce halo mass functions 
that agree both in amplitude and in shape.
  
\subsubsection{Analytic Solution to Parameter Combinations}
    
Instead of searching the parameter space to obtain the right power 
spectrum combinations, we make use of the analytic form of the halo 
mass function.  There are three commonly used analytic formulae for 
halo mass functions.  \citet{press74} developed the first 
theoretical framework based on the spherical collapse model (also 
see \citealt{bond91}) and provided the first theoretical formula 
(PS formula) for the halo mass function, which is still widely used. 
The PS formula is known to underestimate the halo abundance in the 
high mass tail (see \citealt{jenkins01} and references therein). 
\citet{sheth99} give an analytic formula (ST formula) obtained by 
fitting to the results of $N$-body simulations, which turns out to 
be consistent with the ellipsoidal collapse model \citep{sheth01}. 
Based on more simulation results, \citet{jenkins01} provide the most 
accurate fitting formula for halo mass functions.  However, this 
formula includes an absolute value of a function, which makes it 
difficult to use for our present purpose: obtaining an analytic 
solution of parameter combinations that lead to a good match in halo 
mass functions.  We therefore use the slightly less accurate ST 
formula.

In the notation of \citet{jenkins01}, the mass function is defined 
in terms of the dimensionless function
\begin{equation}
f(\nu)\equiv\frac{M}{\rho_0}\frac{{\rm d}n(M)}{{\rm d}\ln \sigma^{-1}},
\end{equation}
and the ST formula for the halo mass function can be expressed as 
\citep{sheth99}
\begin{equation}
f(\nu)=A\sqrt{\frac{2a}{\pi}}\left[1+(a\nu^2)^{-p}\right]\exp(-a\nu^2/2),
~~~ \nu=\delta_c/\sigma(M),
\end{equation}
where $n(M)$ is the abundance of halos with mass less than $M$, 
$\delta_c \approx 1.69$ is the threshold density contrast for 
collapse, $\sigma(M)$ is the r.m.s. fluctuation of the mass density
at a mass scale $M$, $\rho_0$ is the mean density of the current 
universe, $A=0.3222$, $a=0.707$, and $p=0.3$. The mass function can 
be written as
\begin{equation}
\frac{{\rm d}n}{{\rm d}\ln M}=\frac{{\rm d}n}{{\rm d}\ln \sigma^{-1}}
\frac{{\rm d}\ln \sigma^{-1}}{{\rm d}\ln M} =\alpha \frac{\rho_0}{M}f(\nu),
\end{equation}
where $\alpha={\rm d}\ln \sigma^{-1} /{\rm d}\ln M=(3+n_{\rm eff})/6$ 
characterizes the local index $n_{\rm eff}$ of the power spectrum 
and is determined by the shape of the power spectrum. The slope of 
the mass function is
\begin{equation}
\frac{{\rm d}}{{\rm d}\ln M}\left(\frac{{\rm d}n}{{\rm d}\ln M}\right)
=\frac{{\rm d}n}{{\rm d}\ln M}\left(\alpha \frac{{\rm d}\ln f}{{\rm d}\ln \nu}
-1+\frac{{\rm d}\ln \alpha}{{\rm d}\ln M}\right).
\end{equation}
We use a power-law to locally approximate the power spectrum (so that 
the last term of the right-hand side of equation~(8) reduces to zero).

We require that the halo mass function of a given cosmological model 
match as closely as possible that of the central model at some mass 
scale.  In other words, at this mass scale, both the amplitudes 
(eq.~7) and the slopes (eq.~8) of the two halo mass functions should 
be equal. We thus solve these two equations for two unknowns 
($\alpha$,$\nu$) for each cosmological model. All the information 
contained in the power spectrum is fully embedded in ($\alpha$,$\nu$). 
Therefore, by solving for these two parameters at a given mass scale, 
we can also determine the normalization $\sigma_8$ and shape of the 
power spectrum.

The power spectrum can be expressed as 
$P(k) \propto k^{n_s}T^2(k;\Gamma)$, where $n_s$ is the spectral 
index of the inflationary power spectrum and $T(k;\Gamma)$ is the 
transfer function with shape parameter $\Gamma$.  A change in either 
$n_s$ or $\Gamma$ leads to a change in the shape of the power 
spectrum. We first keep $\Gamma$ fixed and only tilt $P(k)$. Since  
$\alpha=(3+n_{\rm eff})/6$ and $n_{\rm eff}$ is simply the sum of 
$n_s$ and an index given by the transfer function, it is 
straightforward to obtain the change in $n_s$ with respect to that 
of the central model: $\Delta n_s = 6\Delta\alpha$. Once the shape 
of $P(k)$ is determined, the normalization $\sigma_8$ can be obtained 
from $\nu$. Assuming that mass functions are matched against that of 
the central model at a cluster mass scale 
$M=5\times10^{14} h^{-1}M_\odot$, we calculate $n_s$ and $\sigma_8$ 
as a function of $\Omega_m$.  The left panel of 
Figure~\ref{fig:case3par} shows that both $n_s$ and $\sigma_8$ 
increase sharply towards low $\Omega_m$ and drop slowly towards high 
$\Omega_m$. In order to match with the halo mass function of the 
central model, we need $n_s=$1.33, 0.82, 0.42 and $\sigma_8=$1.16, 
0.78, 0.55 for models with $\Omega_m=$0.2, 0.4, and 1.0, respectively. 

Alternatively, we can keep the spectral index $n_s$ fixed ($n_s$=1 
in our calculation) and change the shape parameter $\Gamma$.  The 
right panel of Figure~\ref{fig:case3par} shows $\Gamma$ as a function 
of $\Omega_m$ when halo mass functions are matched at 
$M=5\times10^{14} h^{-1}M_\odot$.  The shape parameter $\Gamma$ also 
has a sharp increase towards low $\Omega_m$ and a slow drop towards 
high $\Omega_m$.  For models with $\Omega_m=$0.2, 0.4, and 1.0, 
solutions to $\Gamma$ are 0.35, 0.14, and 0.06, respectively. In this 
case, the normalization $\sigma_8$ of each model is almost identical 
to that of the tilted model, which indicates that the two ways of 
changing the shape of the power spectrum are equivalent at this mass 
scale.  Although pure-$n_s$ and pure-$\Gamma$ changes alter
the power spectrum in different ways, they produce power spectra 
(and thus mass functions) that match closely over a fairly large
dynamic range, covering most of the regime in which $\sigma$ is
of order unity.
The near degeneracy can be broken by examining the low mass 
end of the halo mass functions.
More generally, we could allow both $n_s$ 
and $\Gamma$ to change.  In that case, for each value of $\Omega_m$, 
there would be a 
1-dimensional locus in the $n_s-\Gamma$ plane along which models would 
satisfy the requirement of matching the central model's halo mass 
function at some mass scale. However, we expect that results from this 
more general case would be similar to those of the two simple cases.

\subsubsection{Simulation Results}

Using the power spectrum parameters given by the above analytic 
solutions, we run simulations and make comparisons with the central 
model.  Here we only show results for different tilt models, since 
models with different values of $\Gamma$ give similar results.

Figure~\ref{fig:case3mf} shows cumulative halo mass functions for 
the four cosmological models.  The mass functions agree with each 
other quite well over a large mass range (from 
$\sim 10^{13}h^{-1}M_\odot$ to $\sim 10^{15}h^{-1}M_\odot$), with 
only $\sim 10\%$ relative differences.  We do not expect the 
cumulative mass functions to match exactly both in amplitude and in 
slope at $M=5\times10^{14} h^{-1}M_\odot$, for three reasons.  First, 
the ST formula is not a perfect fit to the halo mass functions
in the simulations. Second, we locally approximate the power 
spectrum as a power-law in order to obtain analytic solutions for 
the power spectrum parameters. Third, we require a match for 
differential mass functions instead of cumulative mass functions. 
Nevertheless, the cumulative mass functions trace each other 
remarkably well. The non-linear mass scales are quite different for 
these four models:
$M_*=3.10 \times 10^{13} h^{-1}M_\odot$, $1.03 \times 10^{13} h^{-1}M_\odot$,
$3.66 \times 10^{12} h^{-1}M_\odot$, and $7.50 \times 10^8 h^{-1}M_\odot$,
for $\Omega_m=0.2$, 0.3, 0.4, 1.0, respectively.

Figure~\ref{fig:case3xi} shows two-point correlation functions of 
the mass and halos. Mass is more strongly clustered in low 
$\Omega_m$ models, having a correlation function that is steeper and 
higher in amplitude than in higher $\Omega_m$ models. However, the 
clustering of halos exhibits the opposite trend. Halos in the 
$\Omega_m=1$ model are highly biased with respect to those of other 
models, as one might guess from the low nonlinear mass of this model, 
and this bias more than compensates for the weaker mass clustering. 
Conversely, at the mass scale $M=1.03\times10^{13}h^{-1}M_\odot$, 
shown in the upper right panel of Figure~\ref{fig:case3xi}, halos in 
the $\Omega_m=0.2$ model are actually anti-biased with respect to 
the mass. Figure~\ref{fig:case2xi} shows that for cluster normalized 
models with fixed $P(k)$ shape, the effects of bias almost exactly cancel 
the change in $\sigma_8$, leaving $\xi_h(r)$ nearly independent of 
$\Omega_m$. Since our mass function matching approach imposes a 
redder $P(k)$ shape (lower $n_s$ or $\Gamma$) for higher $\Omega_m$, 
and increased large scale power amplifies the importance of bias, it 
is not surprising that bias {\it over} compensates mass clustering 
changes in these models, leaving $\xi_h(r)$ steadily dependent on 
$\Omega_m$.

Figure~\ref{fig:case3ve} shows the mean pairwise radial velocity and 
pairwise velocity dispersion for the four cosmological models. As in
the case of cluster normalized models, models with higher $\Omega_m$ 
have larger mean pairwise velocities and velocity dispersions than
lower $\Omega_m$ models.  The dependence on $\Omega_m$ is stronger 
and more systematic than that in Figure~\ref{fig:case2ve}, again a 
sign of the redder power spectrum of higher $\Omega_m$ models.
 
\section{Summary and Discussion}
\label{sec:discussion}

We have examined how changes to the matter density parameter 
$\Omega_m$ and the shape and amplitude of the linear power spectrum 
$P(k)$ affect the halo mass function, the two-point halo correlation 
function, and the first and second moments of the halo pairwise 
velocity distribution.  A change in $\Omega_m$, at fixed $P(k)$, 
simply shifts the halo mass scale.  Therefore, if halo masses are 
scaled in proportion to $\Omega_m$, halo populations of different 
$\Omega_m$ models have identical mass functions and clustering 
properties.  However, mean pairwise velocities and pairwise 
velocity dispersions, which scale 
as $\Omega_m^{0.6}$ on large scales (and fairly far into the 
non-linear regime) break this degeneracy.  A change of the power 
spectrum amplitude ($\sigma_8$) along with $\Omega_m$, maintaining 
a cluster normalization condition $\sigma_8 \propto \Omega_m^{-0.5}$, 
produces halo populations that have nearly identical spatial 
clustering at a fixed physical mass scale and systematic but small 
differences in the mean pairwise velocities and pairwise velocity 
dispersions. However, while these cluster normalized models have the 
same halo space density at $M \sim$ several $\times 10^{14} M_\odot$, 
the shapes of their mass functions are systematically different. If 
we further allow the shape of the power spectrum to change (either 
by tilting it or by changing its shape parameter $\Gamma$), we can 
produce halo populations whose mass functions match very well both 
in amplitude and in shape, over a large mass range. However, the 
changes to the power spectrum shape cause correlation functions, 
mean pairwise velocities, and velocity dispersions for halos in the 
same mass range to differ.  We conclude that the halo populations 
produced by distinct cosmological models are not degenerate. If they 
are indistinguishable by one statistic, they can be told apart using 
another statistic. 

Our results imply that a perfect knowledge of the halo population 
and its properties would allow us to pin down the underlying 
cosmological parameters -- specifically the value of $\Omega_m$ and 
the shape and amplitude of $P(k)$ -- even before bringing in 
constraints from other cosmological measurements.
However, a galaxy redshift survey detects galaxies rather than 
halos. Could changes to the halo occupation distribution (HOD) mask 
the differences in the halo populations of different cosmological 
models?  Models that differ only in $\Omega_m$ lead to the same halo 
mass function and halo clustering in terms of $M/M_*$, but with 
$M_* \propto \Omega_m$.  Their halos could be thus be populated the 
same way as a function of $M/M_*$ to produce indistinguishable galaxy 
spatial clustering.  However, any dynamically sensitive statistics 
-- virial masses of clusters and groups, pairwise velocity dispersions, 
the parameter combination $\Omega_m^{0.6}/b$ inferred from 
redshift-space distortions, the galaxy-mass correlation function from 
galaxy-galaxy lensing -- would distinguish models with different 
values of $\Omega_m$ immediately.  Velocity bias within halos could 
mask some of these changes, but not all of them \citep{berlind02}. 
Cluster-normalized models with different $\Omega_m$ and the same 
$P(k)$ shape have different halo mass functions, and they would 
therefore require a different $P(N|M)$ in order to keep the galaxy 
density fixed.  This change would likely cause differences in the
galaxy clustering even though the halo clustering is similar, since 
galaxy clustering is highly sensitive to $P(N|M)$ (\citealt{berlind02} 
and references therein).  Changing the shape of $P(k)$ by the amount 
required to match mass function shapes makes substantial differences 
to the halo clustering and velocities, which seem unlikely to be 
masked by a change in the HOD. 

These speculations suggest that distinct cosmological models, which 
produce nondegenerate halo populations, cannot have galaxy populations 
that are indistinguishable in every spatial and velocity statistic, 
even if one allows complete freedom in the way that galaxies occupy 
these halos.  We reserve a detailed investigation of this issue for 
future work.  If our optimistic speculation holds true, then high 
precision measurements of galaxy clustering and galaxy-galaxy lensing 
should impose strong constrains on cosmological models without 
reliance on {\it a priori} models of galaxy bias.

\acknowledgments

We thank Volker Springel for advice on GADGET and Changbom Park for 
the use of his PM code. DW thanks the Institute for Advanced Study
for its hospitality and the Ambrose Monell Foundation for its financial 
support during the final stages of this work. This work was supported 
by NSF grant AST-0098584.

\clearpage

\begin{figure}[h]
\caption[pgslice]{\label{fig:pgslice}
Comparison of PM and GADGET simulations evolved from the same initial 
conditions.  Top panels show slices through the particle distributions,
$100\hmpc$ on a side and $10\hmpc$ thick.  Bottom panels
show cumulative halo mass functions; in the right panel these
are divided by the \cite{jenkins01} analytic mass function to allow
more detailed comparison. The minimum group mass plotted corresponds 
to 10 simulation particles.
}
\end{figure}

\begin{figure}[h]
\plotone{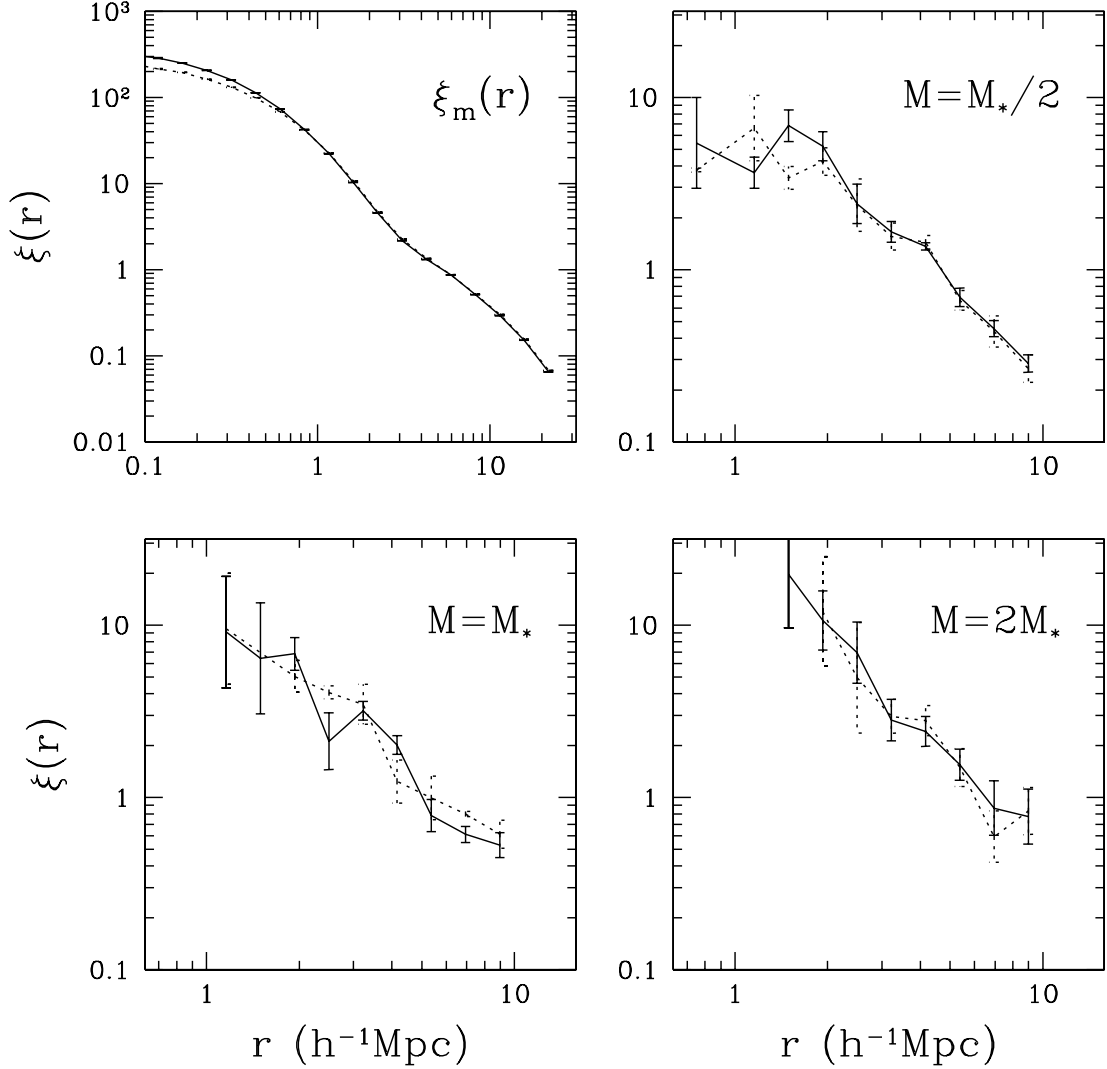}
\caption[pgxicmp]{\label{fig:pgxicmp}
Matter and halo correlation functions of the PM (dotted lines) and
GADGET (solid lines) simulations shown in Figure~\ref{fig:pgslice}.
The upper left panel shows the matter correlation functions and 
remaining panels show halo correlation functions for three 
different mass ranges, centered on $M_*/2$, $M_*$, and $2M_*$, 
where $M_*=1.03\times10^{13}h^{-1}M_\odot$ is the characteristic 
nonlinear mass (corresponding to 123 particles in these 
simulations).  Results are averaged over four realizations, and 
error bars show the run-to-run dispersion divided by 
$\sqrt{N-1}=\sqrt{3}$ to yield the uncertainty in the mean.
}
\end{figure}

\begin{figure}[h]
\plotone{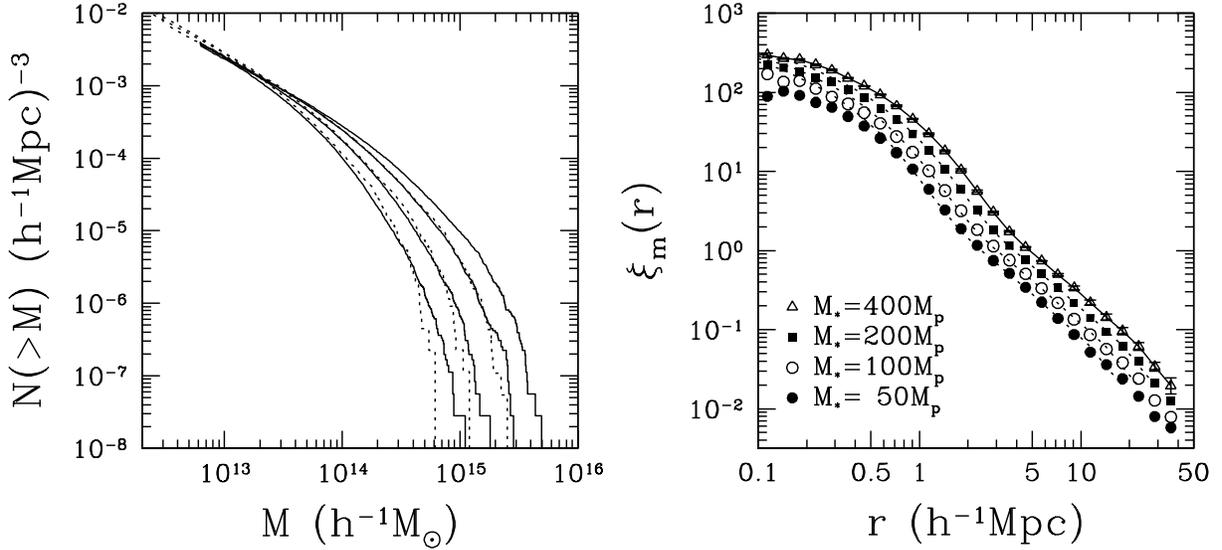}
\caption[scalemf]{\label{fig:scalemf}
Test of self-similar scaling in PM simulations with scale-free 
initial conditions.  Solid lines in the left panel show halo mass 
functions at four evolutionary stages, when $M_*$=50, 100, 200, 
and 400 particles (left to right).  Dotted lines show the 
rightmost solid line scaled  according to 
equation~(\ref{eqn:mscaling}).  Points in the right panel show 
two-point correlation functions of the mass at these four epochs. 
Dotted lines are derived by scaling the mass correlation function 
of the final output, when $M_*=400M_p$, according to 
equation~(\ref{eqn:xiscaling}).  Results are averaged over four 
simulations, and error bars on $\xi(r)$ show the dispersion among 
four independent runs, divided by $\sqrt{3}$. For clarity, error bars
are plotted only for the stage of $M_*=400M_p$, and error bars for 
other stages have similar magnitude. Physical scales 
are assigned by treating the final output as $z=0$ and adopting 
$\sigma_8=1.0$, which implies a comoving size 
$L_{\rm box}=207.62h^{-1}M_\odot$ of the simulation cube and a
total mass $M=2.49\times 10^{18} h^{-1}M_\odot$ in the cube.
}
\end{figure}

\begin{figure}[h]
\plotone{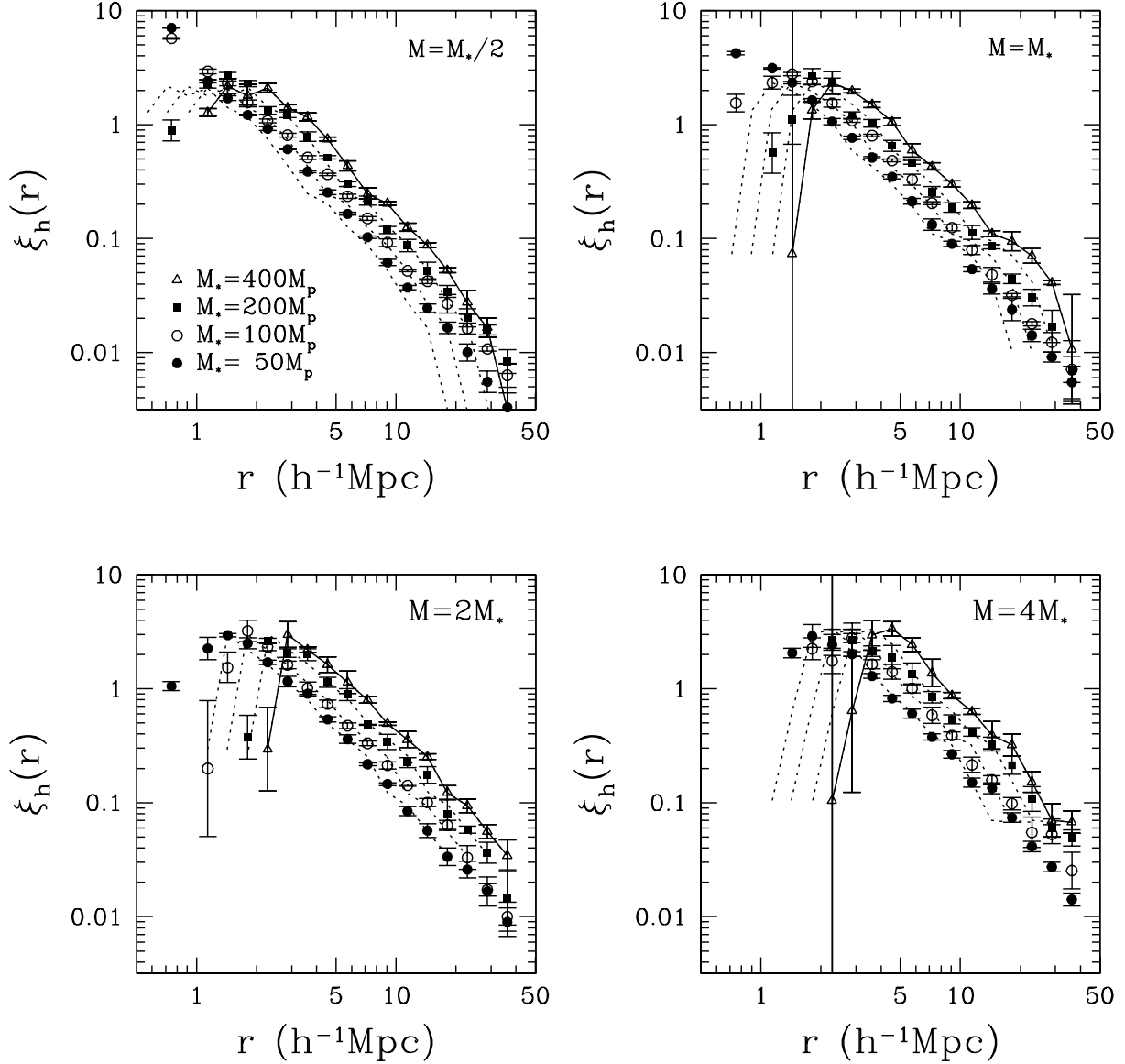}
\caption[scalexi]{\label{fig:scalexi}
Test of self-similar scaling of the halo correlation functions, 
in mass ranges centered on $M_*/2$, $M_*$, $2M_*$, and $4M_*$, as 
indicated.  Points show halo correlation functions at the four
evolutionary stages when $M_*=50$, 100, 200, and 400 particles.
Dotted lines are obtained by scaling the result of the $M_*=400$ 
output (solid line) to earlier stages, according to 
equation~(\ref{eqn:xiscaling}).  Error bars show the dispersion 
among four independent runs, divided by $\sqrt{3}$.
}
\end{figure}

\begin{figure}[h]
\plotone{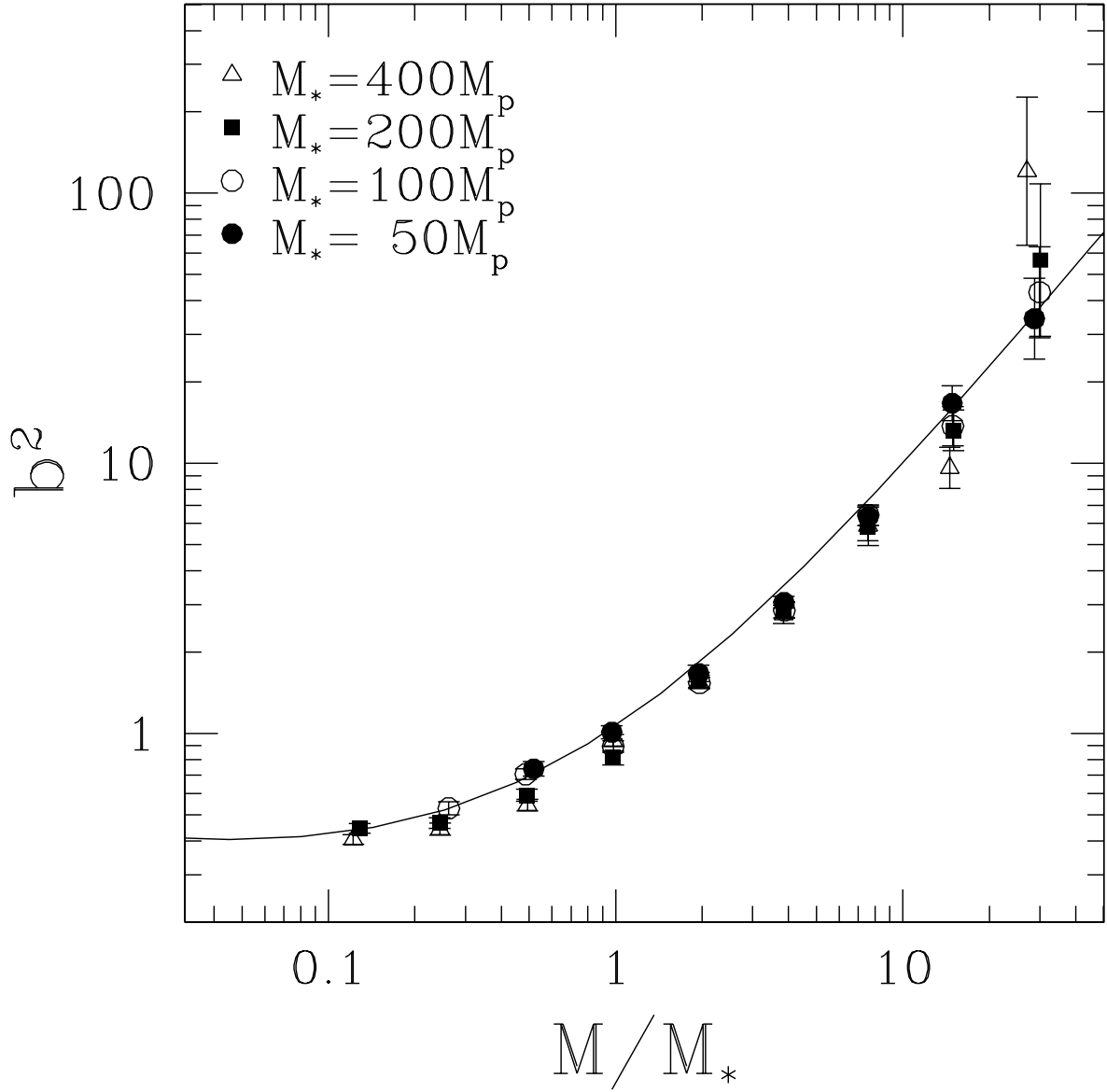}
\caption[scalebias]{\label{fig:scalebias}
Halo bias factor as a function of $M/M_*$ in the scale-free 
simulations.  Points show $b^2$ at the four evolutionary stages
when $M_*=50$, 100, 200, and 400 particles.  The solid line is
calculated using Jing's (\citeyear{jing98}) formula.
}
\end{figure}

\begin{figure}[h]
\epsscale{0.75}
\epsscale{1.0}
\caption[slice]{\label{fig:slice}
Insensitivity of dark matter clustering to the values of 
$\Omega_m$ and $\Omega_\Lambda$.  The top and middle panels show 
slices of particle distributions at $z=0$ from simulations that 
have identical linear theory $P(k)$ and Fourier phases but different 
combinations of $\Omega_m$ and $\Omega_\Lambda$, as indicated.  
The two bottom panels are slices for models (1.0,0.0) and (0.2,0.8) 
at redshifts corresponding to a linear growth factor of 0.5. All 
slices have the size of $100\hmpc \times 100\hmpc \times 10\hmpc$. 
All simulations were performed with GADGET using $100^3$
particles in a $(100\hmpc)^3$ volume and a force resolution
of $\epsilon=0.3\hmpc$.
}
\end{figure}

\begin{figure}[h]
\plotone{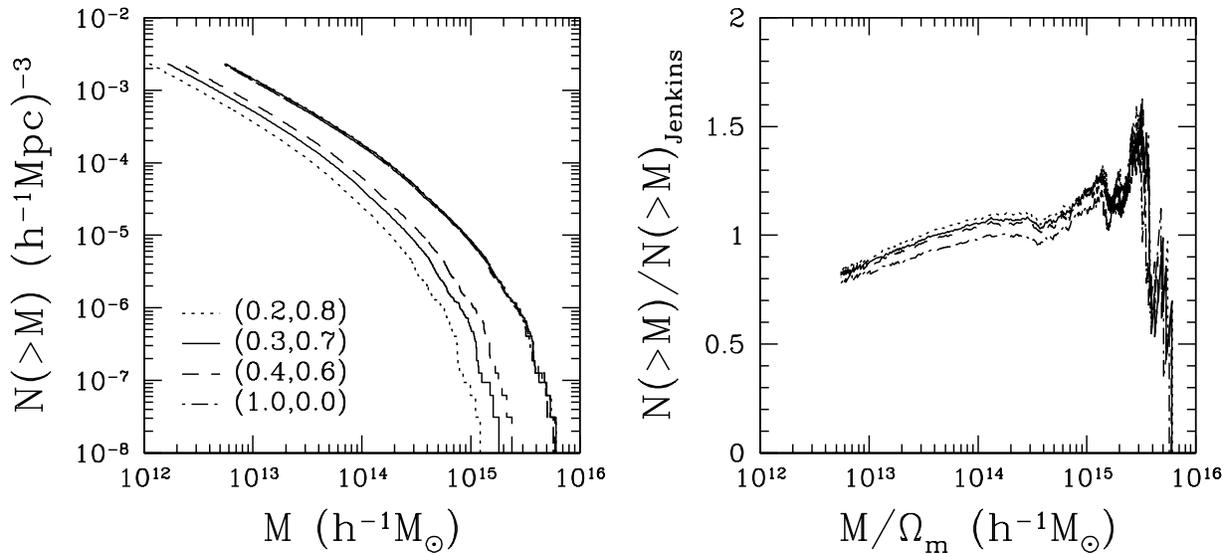}
\caption[case1mf]{\label{fig:case1mf}
Influence of $\Omega_m$ on the halo mass function.  The left panel 
shows cumulative halo mass functions from simulations with four 
different combinations of $(\Omega_m,\Omega_\Lambda)$ as indicated.  
A second curve for each low-$\Omega_m$ model shows the result of 
dividing halo masses by $\Omega_m$ before computing $N(>M)$.  These 
scaled curves are almost perfectly superposed on the solid curve 
representing the $\Omega_m=1.0$ model, demonstrating that a change 
to $\Omega_m$ alone simply shifts the mass scale of the halo mass 
function.  The right panel allows closer inspection of this result,
suppressing dynamic range by dividing each mass function by the
analytic fitting formula of \cite{jenkins01} computed for the 
parameters of the central model $(\Omega_m=0.3,\Omega_\Lambda=0.7)$.
}
\end{figure}

\begin{figure}[h]
\plotone{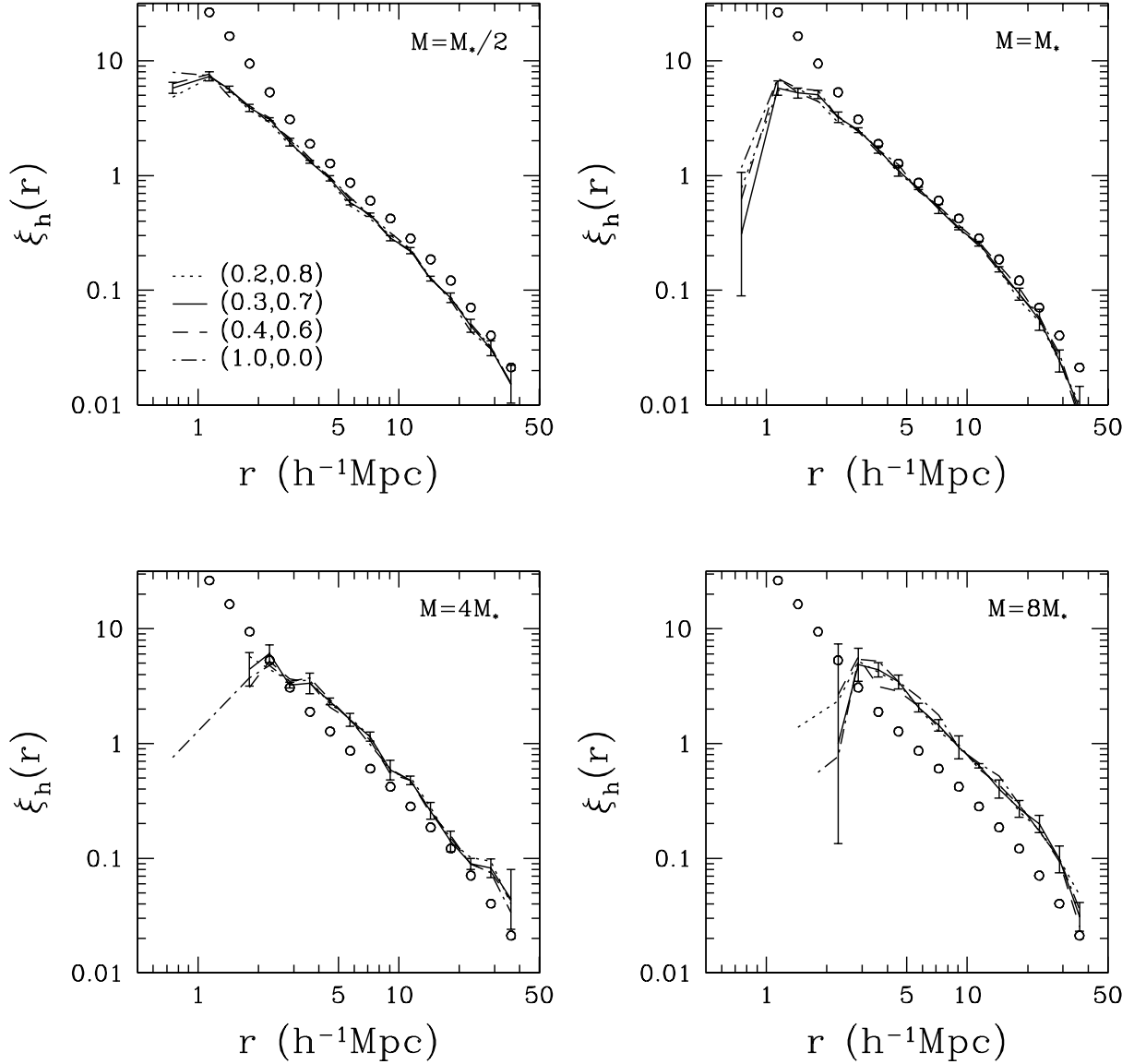}
\caption[case1xi]{\label{fig:case1xi}
Influence of $\Omega_m$ on the halo correlation function, for halos 
in mass ranges centered on $M_*/2$, $M_*$, $4M_*$, and $8M_*$ as 
indicated.  Curves in each panel, nearly superposed, show $\xi_h(r)$ 
for the four indicated combinations of $(\Omega_m,\Omega_\Lambda)$.  
To preserve visual clarity, we plot error bars only on the central 
(0.3,0.7) model, but error bars for other models have similar 
magnitude.  In each panel, open circles show the mass correlation 
function of the central model.
}
\end{figure}

\begin{figure}[h]
\plotone{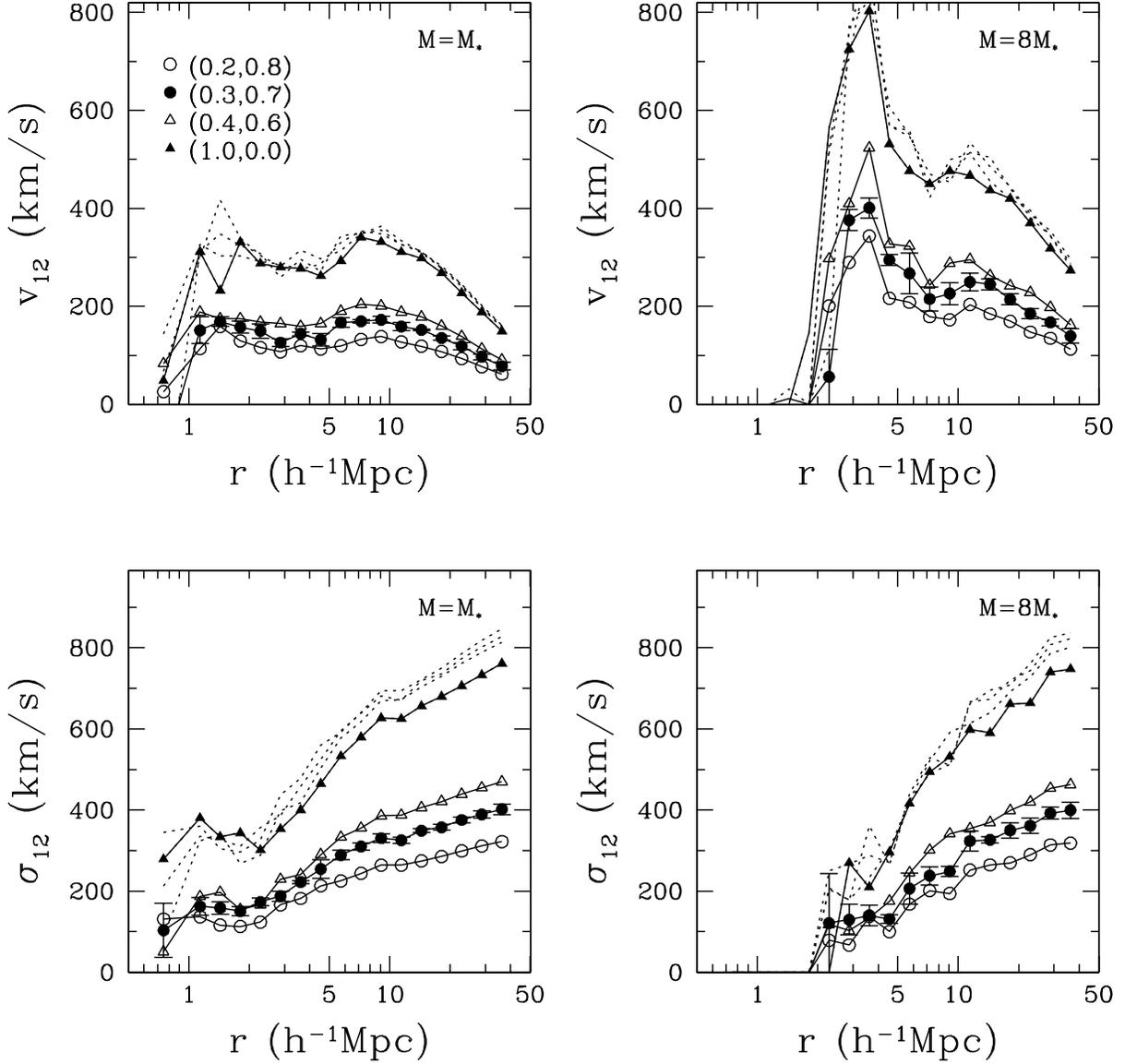}
\caption[case1ve]{\label{fig:case1ve}
Influence of $\Omega_m$ on the mean pairwise velocity (top) and the 
pairwise velocity dispersion (bottom) of halos with mass 
$M\sim M_*$ (left) and $M\sim 8M_*$ (right).  Error bars are plotted 
only for the central model; they increase with $\Omega_m$ and are 
about twice as large for $\Omega_m=1$ as for $\Omega_m=0.2$.  Dotted 
curves are obtained by dividing the velocity by $\Omega_m^{0.6}$;
their agreement with the $\Omega_m=1$ curve demonstrates that the 
$\Omega_m$ influence is captured almost entirely by the linear 
theory velocity scaling.
} 
\end{figure}

\begin{figure}[h]
\plotone{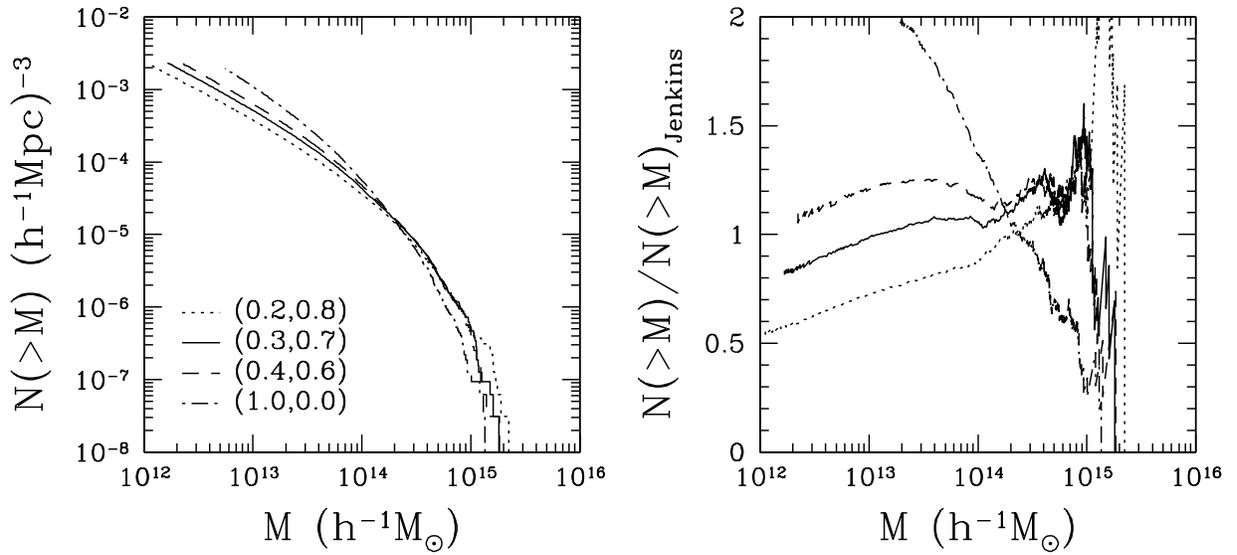}
\caption[case2mf]{\label{fig:case2mf}
Halo mass functions for cluster normalized models, in which the 
shape of $P(k)$ is fixed but the amplitude is scaled by 
$\sigma_8\propto \Omega_m^{-0.5}$.
The left panel shows halo mass functions from simulations with 
four different $(\Omega_m,\Omega_\Lambda)$, as indicated.
In the right panel, these mass functions are divided by the
\cite{jenkins01} fitting formula for the central (0.3,0.7) model.
}
\end{figure}

\begin{figure}[h]
\plotone{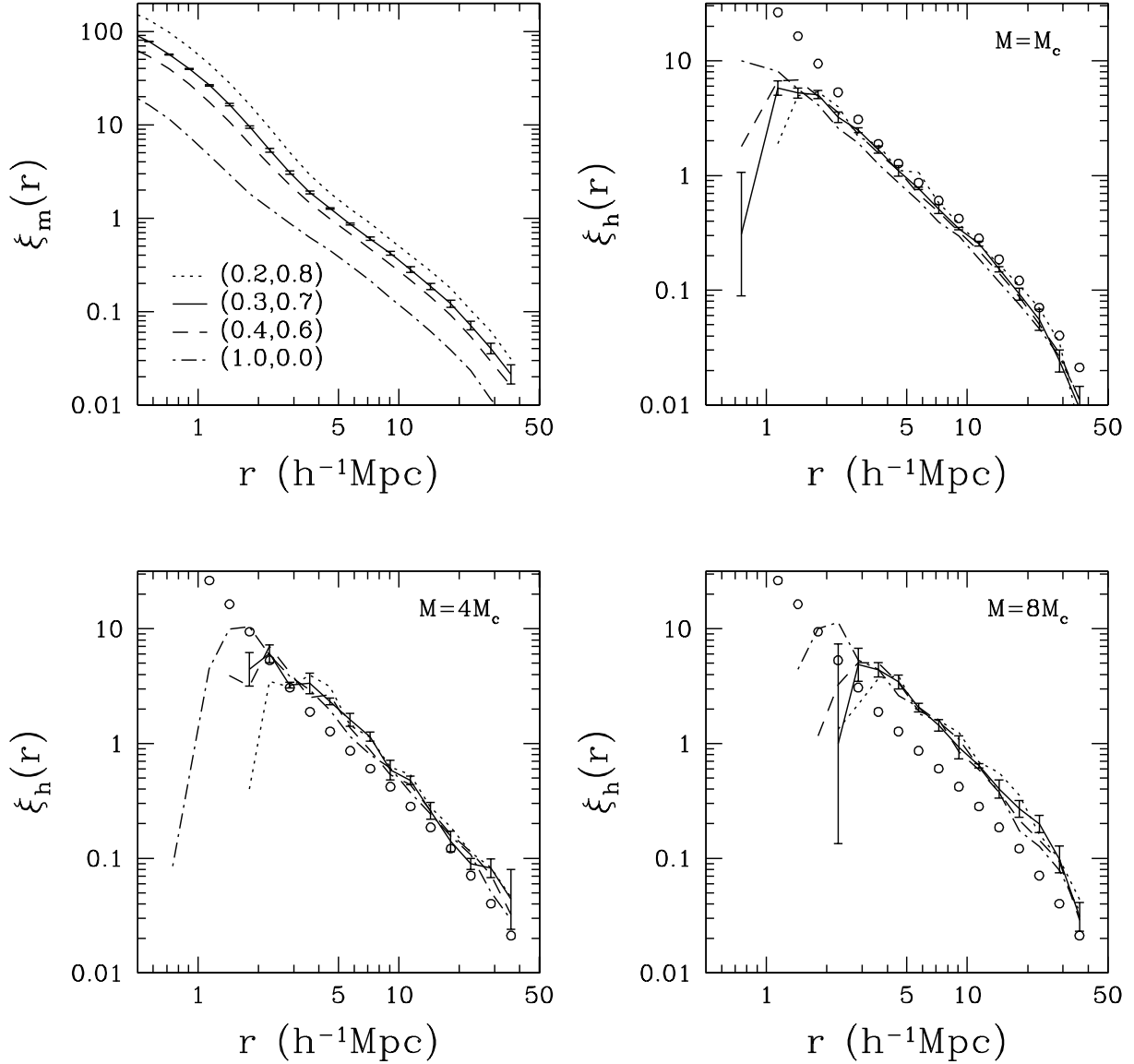}
\caption[case2xi]{\label{fig:case2xi}
Mass and halo correlation functions for cluster normalized models.
The upper left panel shows the mass correlation function.
The remaining three panels show correlation functions of halos in
mass ranges centered on $M_c$, $4M_c$, and $8M_c$,
where $M_c=1.03\times10^{13}h^{-1}M_\odot$ 
corresponds to $M_*$ of the central (0.3,0.7) model.  Open circles 
show the mass correlation function for the central model.  Error 
bars are plotted for the central model and have similar magnitude
for other models.
} 
\end{figure}

\begin{figure}[h]
\plotone{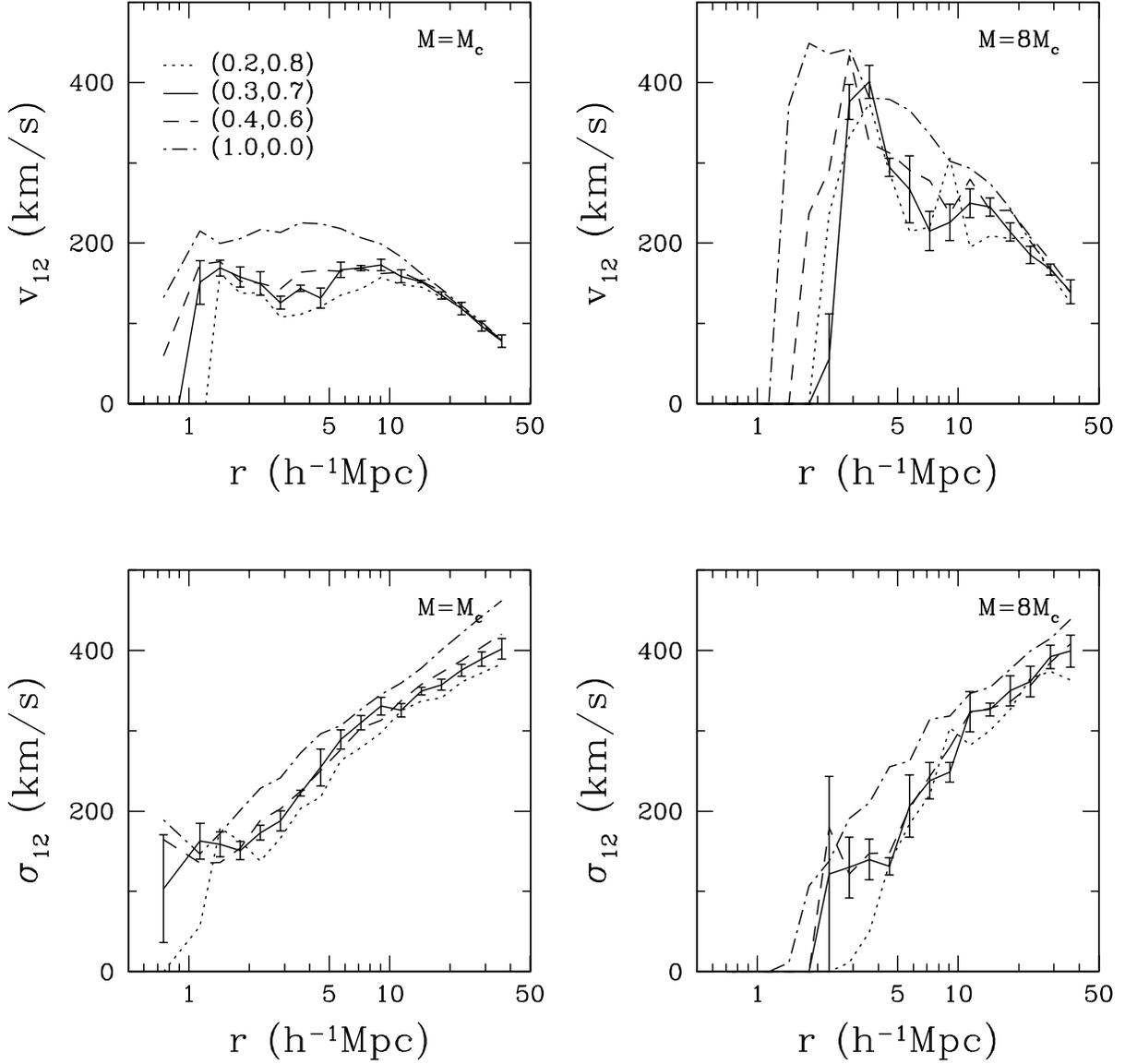}
\caption[case2ve]{\label{fig:case2ve}
Halo velocity statistics for cluster normalized models.  Top panels 
show the mean pairwise radial velocities for halos with $M \sim M_c$ 
(left) and $M \sim 8M_c$ (right), where 
$M_c=1.03\times10^{13}h^{-1}M_\odot$ corresponds to $M_*$ of the 
central model.  Bottom panels show pairwise radial velocity 
dispersions.  Error bars are plotted for the central model and have 
similar magnitude for other models.
}
\end{figure}

\begin{figure}[h]
\plotone{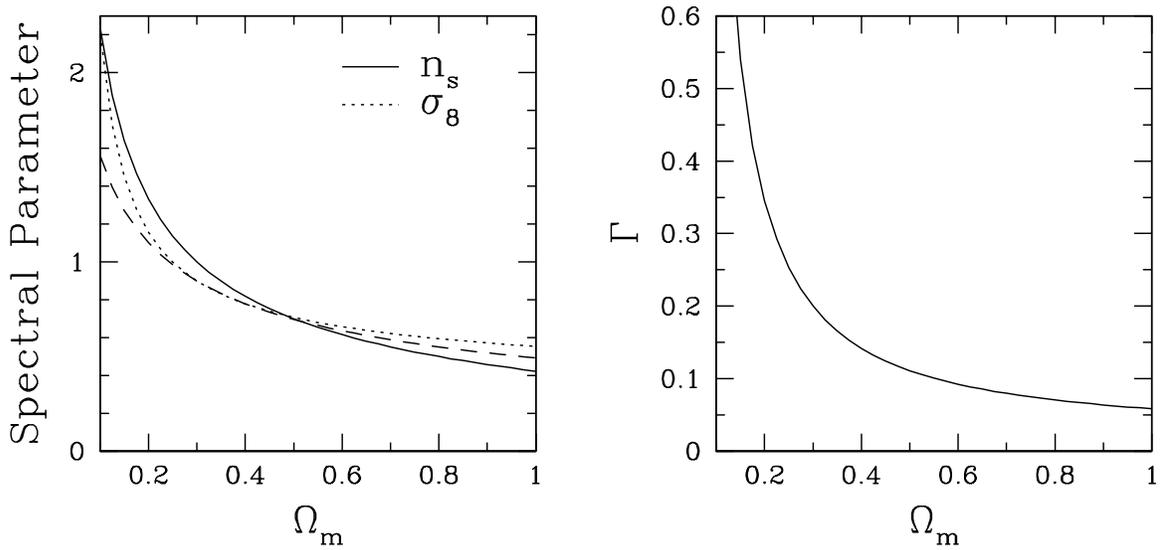}
\caption[case3par]{\label{fig:case3par}
Changes in power spectrum parameters required to match the 
amplitude and slope of the halo mass function at 
$M=5\times10^{14}h^{-1}M_\odot$.  The left panel shows the values of 
$\sigma_8$ (dotted line) and $n_s$ (solid line) required to match 
the mass function of a model with 
$\Omega_m=0.3$, $\sigma_8=0.9$, $n_s=1.0$, and $\Gamma=0.2$, 
when $\Gamma$ is fixed at 0.2.  The values of $\sigma_8$ we adopt for
cluster-normalized models 
with fixed $P(k)$ shape
are plotted as a dashed line, for
comparison.  The right panel shows the required 
value of $\Gamma$ if $n_s$ is held fixed at 1.0; the corresponding 
values of $\sigma_8$ are nearly identical to those in the left 
panel.  Curves are calculated using the Sheth-Tormen 
(\citeyear{sheth99}) analytic mass function, as discussed in the 
text.
}
\end{figure}

\begin{figure}[h]
\plotone{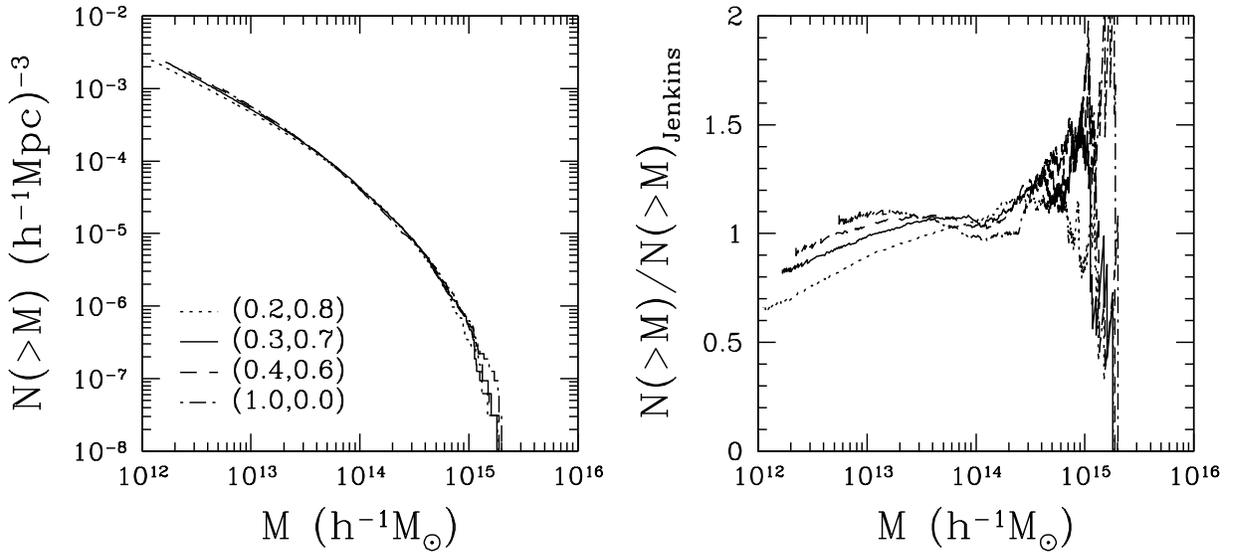}
\caption[case3mf]{\label{fig:case3mf}
Halo mass functions for models that have been matched at 
$M=5\times10^{14}h^{-1}M_\odot$ by changing $n_s$ and $\sigma_8$. 
As expected, the change in $n_s$ yields much better agreement of 
the mass functions than found for cluster normalized models with no
change to the shape of $P(k)$ (Figure~\ref{fig:case2mf}).
}
\end{figure}

\begin{figure}[h]
\plotone{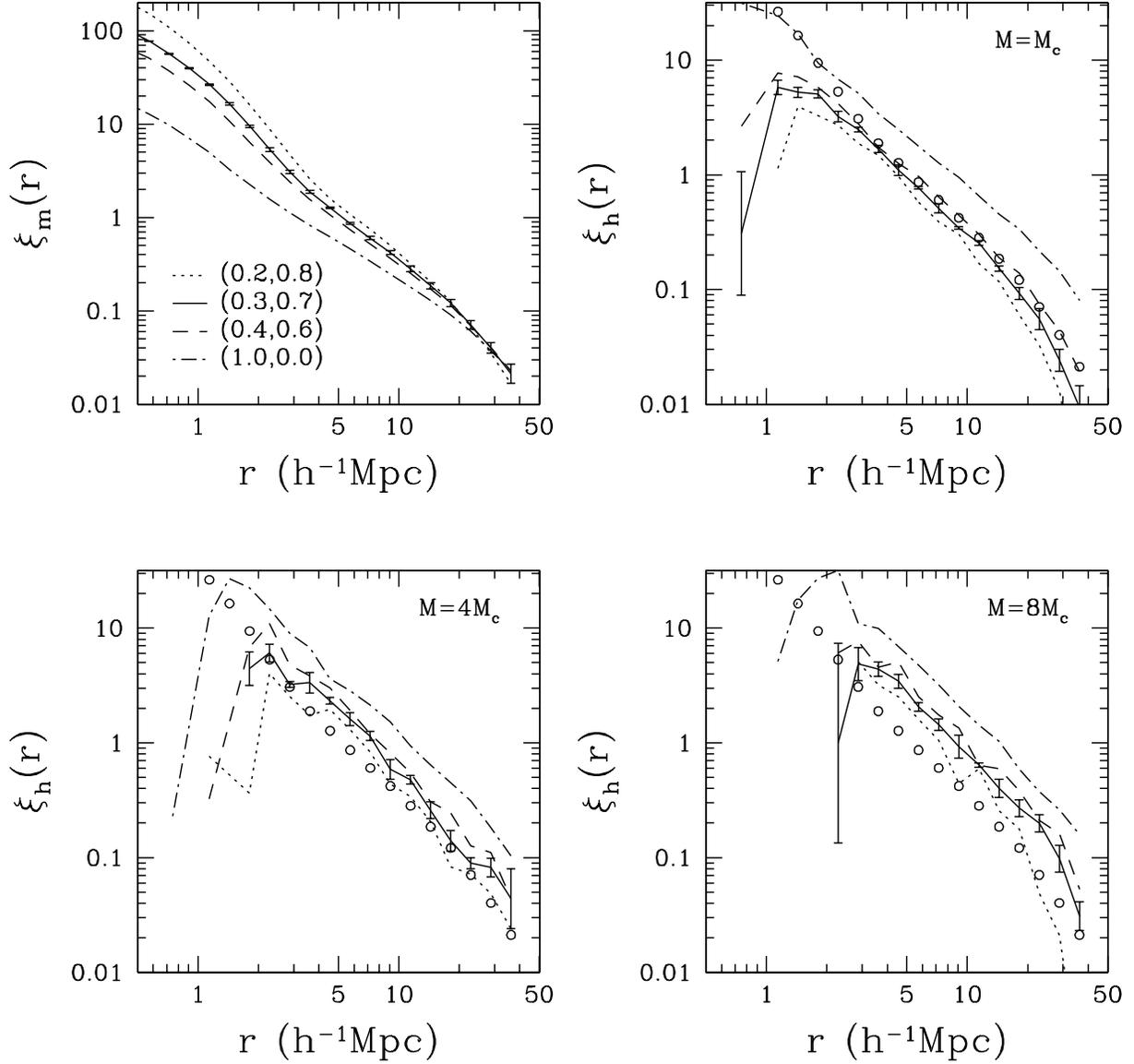}
\caption[case3xi]{\label{fig:case3xi}
Mass and halo correlation functions for the four models of 
Figure~\ref{fig:case3mf}, in the same format as Figure~\ref{fig:case2xi}.
Halo mass ranges are centered on $M_c$, $4M_c$, $8M_c$,
where $M_c=1.03\times10^{13}h^{-1}M_\odot$ corresponds to $M_*$ of 
the central model. Open circles in these three panels show the mass 
correlation function of this model.  Error bars are plotted for the 
central model and have similar magnitude for other models.  
Changing $n_s$ to match the shapes of the halo mass functions leads 
to substantial differences in halo clustering, in contrast to 
Figure~\ref{fig:case2xi}.
}
\end{figure}

\begin{figure}[h]
\plotone{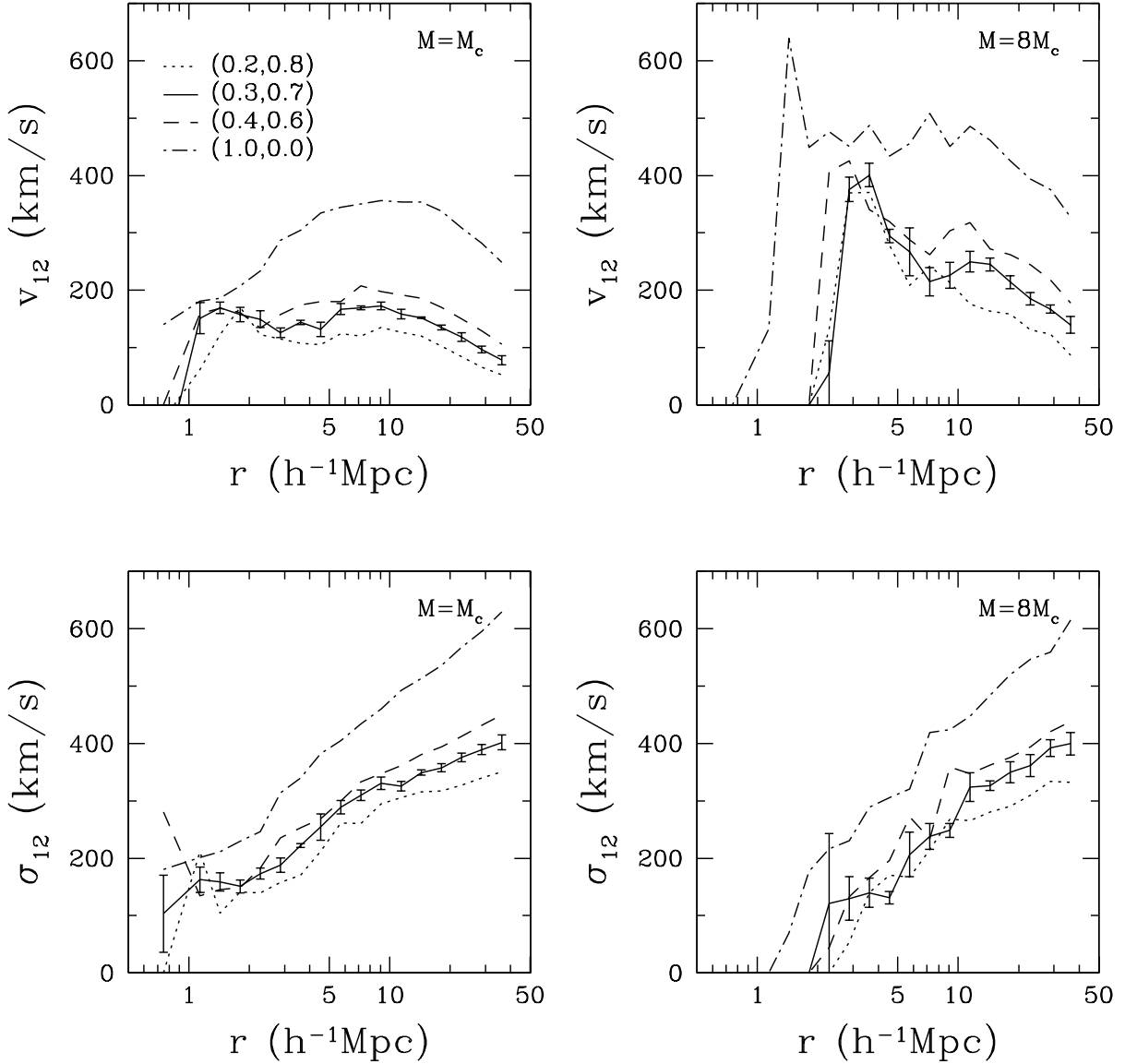}
\caption[case3ve]{\label{fig:case3ve}
Halo velocity statistics for the four models shown in 
Figures~\ref{fig:case3mf} and~\ref{fig:case3xi}, in the same format
as Figure~\ref{fig:case2ve}.  Error bars are plotted for the 
central model; they increase with $\Omega_m$ and are about twice
as large for $\Omega_m=1$ as for $\Omega_m=0.2$.  Changing $n_s$ 
to match the halo mass function shapes leads to significant 
systematic differences in the mean pairwise velocities and pairwise 
velocity dispersions as a function of $\Omega_m$, larger than those 
in Figure~\ref{fig:case2ve}.
}
\end{figure}

\end{document}